\documentclass[aps,prb,twocolumn,epsf,float,letter]{revtex4}
\usepackage{graphicx}
\usepackage{amsmath,amsthm,amssymb,mathrsfs}
\usepackage{bm}
\usepackage{setspace}
\begin{document}

\title{Spin transfer torques generated by the anomalous Hall effect
  and anisotropic magnetoresistance.}  
\author{Tomohiro Taniguchi}
\affiliation{National Institute of Advanced Industrial Science and Technology
    (AIST), Spintronics Research Center, Tsukuba, Ibaraki 305-8568,
    Japan}
\affiliation{Center for Nanoscale Science and Technology, National
Institute of Standards and Technology, Gaithersburg, Maryland
20899-6202, USA }
\author{J. Grollier}
\affiliation{Unit\'e Mixte de Physique CNRS/Thales and Universit\'e
  Paris Sud 11, 1 Avenue Fresnel, 91767 Palaiseau, France}  
\author{M. D. Stiles}
\affiliation{Center for Nanoscale Science and Technology, National
Institute of Standards and Technology, Gaithersburg, Maryland
20899-6202, USA }

\begin{abstract}
  Spin-orbit coupling in ferromagnets gives rise to the anomalous Hall
  effect and the anisotropic magnetoresistance, both of which can be
  used to create spin-transfer torques in a similar manner as the spin
  Hall effect.  In this paper we show how these effects can be used to
  reliably switch perpendicularly magnetized layers and to move domain
  walls.  A drift-diffusion treatment of the anomalous Hall effect and
  the anisotropic magnetoresistance describes the spin currents that
  flow in directions perpendicular to the electric field. In systems
  with two ferromagnetic layers separated by a spacer layer, an
  in-plane electric field cause spin currents to be injected from one
  layer into the other, creating spin transfer torques. Unlike the
  related spin Hall effect in non-magnetic materials, the anomalous
  Hall effect and the anisotropic magnetoresistance allow control of
  the orientation of the injected spins, and hence torques, by
  changing the direction of the magnetization in the injecting
  layer. The torques on one layer show a rich angular dependence as a
  function of the orientation of the magnetization in the other
  layer. The control of the torques afforded by changing the
  orientation of the magnetization in a fixed layer makes it possible
  to reliably switch a perpendicularly magnetized free layer.  Our
  calculated critical current densities for a representative
  CoFe/Cu/FePt structure show that the switching can be efficient for
  appropriate material choices.  Similarly, control of the
  magnetization direction can drive domain wall motion, as shown for
  NiFe/Cu/NiFe structures.
\end{abstract}

\maketitle


\section{Introduction}
\label{sec:intro}

The use of spin-orbit coupling to generate spin-transfer
torques\cite{Berger:1984,Slonczewski:1996,Berger:1996,Stiles:2006,Ralph:2007}
raises the possibility of new types of devices and more efficient
versions of existing devices.  In general, the spin-orbit coupling in
these studies has
been provided by a non-magnetic heavy metal layer such as Pt.
Here, we show that replacing
this non-magnetic layer by a
ferromagnetic layer and a thin spacer layer offers potential
advantages in device design.  In existing approaches, spin-orbit
torques\cite{Ando:2008,Liu:2011} typically derive from the spin Hall
effect\cite{Dyakonov:1971,Hirsch:1999,Zhang:2000} in the bulk of
non-magnetic layers or from spin-orbit torques localized at
the interface between such a layer and a ferromagnetic
layer.\cite{Bychkov:1984,Edelstein:1990,Obata:2008,Manchon:2008,MatosAbiague:2009,Wang:2012,Kim:2012,Pesin:2012,Bijl:2012}
The resulting torques may lead to more efficient switching of memory
elements\cite{Miron:2011a,Liu:2012a,Garello:2013a,Lee:2013,Lee:2014}
or domain wall
motion.\cite{Miron:2011b,Haazen:2013,Emori:2013,Ryu:2013,Yoshimura:2014,Seo:2012,Thiaville:2012}
Considerable
experimental\cite{KimJ:2013,Qiu:2013,Fan:2014,Liu:2014,Garello:2013,Qiu:2014}
and
theoretical\cite{Haney:2013a,Haney:2013b,Freimuth:2013,Freimuth:2014,Kurebayashi:2014}
work has been devoted to characterizing these torques so as to
understand the details of their origin.  However, device design
possibilities based on heavy metal layers are somewhat limited by the
fact that the form of the torques is determined by the geometry of the
device, that is, the direction of the current flow and the interface
normal.  We show that replacing the non-magnetic heavy metal by a
ferromagnetic layer and a thin spacer layer gives greater control
over the form of the torque because it is controlled by the direction
of the magnetization, which can be varied, rather than the geometry.

Historically, the earliest spintronic effects, discovered before the
electron was known to have a spin, were the anisotropic
magnetoresistance,\cite{Thomson:1857,McGuire:1975} and the anomalous
Hall
effect.\cite{Kundt:1893,Pugh:1953,Karplus:1954,Sinitsyn:2008,Nagaosa:2010}
Both of these effects are caused by spin-orbit coupling, but because
of the strong coupling between spin currents and charge currents in
ferromagnets, these are typically discussed in terms of the resulting
charge currents and voltages. Very recently, several
groups\cite{Miao:2013,Azevedo:2014,Wang:2014,Tsukahara:2014,Weiler:2014}
measured what they described as the inverse spin 
Hall effect in permalloy, a nickel-iron alloy.  This result raises the
point that a spin current will always accompany the charge current
caused by the anomalous Hall effect\cite{Zhang:2000} and the spin
current will vary with the angle between the magnetization and the
charge current as in the anisotropic magnetoresistance.  We show that both the
anomalous Hall effect and anisotropic magnetoresistance can be
exploited to generate spin currents and spin transfer torques in much
the same way as the spin Hall effect.

The spin Hall effect\cite{Dyakonov:1971,Hirsch:1999,Zhang:2000} occurs
in metals, particularly heavy metals with strong spin-orbit coupling.
When an electric field is applied in a particular direction, a spin
current flows in all directions perpendicular to the field with spins
oriented perpendicularly to their flow. That is, for an electric field
in the $\hat{\bf E}$ direction, there is a spin current in every
direction ${\bf e}$ perpendicular to the electric field $\hat{\bf
  e}\cdot\hat{\bf E}=0$ with spins pointing in the $\hat{\bf
  e}\times\hat{\bf E}$ direction.  This spin current can be written in
the form $Q_{ij} = (-\hbar/2e) \sigma_{\rm SH} \epsilon_{ijk} E_k$, where the second index of
the tensor spin current ${\bf Q}$ refers to the real space direction
of flow and the first index refers to the orientation of the spin that
is flowing.  ${\bf E}$ is the electric field, $\sigma_{\rm SH}$ is the
spin Hall conductivity, and $\epsilon_{ijk}$ is
the Levi-Civita symbol.  Repeated indices (here $k$) are summed over
(here summing over $k=\ x,\ y,\ z$).  The spin current arises through
either intrinsic mechanisms,\cite{Guo:2008,Tanaka:2008} that is
through the spin-orbit coupling in the band structure, or extrinsic
mechanisms\cite{Gradhand:2010,Lowitzer:2011} through the spin-orbit
coupling in the impurity scattering.

The same spin-orbit effects occur in ferromagnets but are complicated
by the exchange potential that gives rise to spin split band
structures and spin-dependent conductivities.  One complication is
that in a ferromagnet any spin that is
transverse to the magnetization precesses rapidly, so any transverse
spin accumulation or spin current dephases quickly due to this
precession.  Thus, it becomes a very good approximation to treat the
spins in a ferromagnet as parallel or antiparallel to the
magnetization.  Then, the tensor spin current in a ferromagnet has
spins pointing in the direction of the magnetization ${\bf m}$ flowing
in the ${\bf j}_{\rm s}$ direction, or ${\bf Q} \sim {\bf m} \otimes
{\bf j}_{\rm s}$.  This feature plays a crucial role in the results
below. It allows control of the direction of the spins injected
into other layers due to spin-orbit effects simply by changing ${\bf
  m}$. Such control does not exist with the spin Hall effect where the
direction the spins point when injected into another layers is ${\bf
  n}\times {\bf E}$, where ${\bf n}$ is the interface normal
direction.

A second complication is that majority and minority electrons see very
different potentials so the spin-orbit scattering that gives rise to
pure spin currents in non-magnets gives rise to a charge current as
well as a spin current.  This charge current is the current measured
in the anomalous Hall effect, whose direction is given by $\mathbf{m}
\times \mathbf{E}$.  Therefore, the spin current excited by the
anomalous Hall effect has spins pointing the ${\bf m}$ direction
flowing in the $\mathbf{m} \times \mathbf{E}$ direction, that is 
\begin{eqnarray}
  {\bf Q}
&=& \frac{-\hbar}{2e} \zeta \sigma_{\rm AH} \mathbf{m} \otimes \mathbf{m} \times \mathbf{E}
\nonumber\\
Q_{ij}&=& \frac{-\hbar}{2e} \zeta \sigma_{\rm AH} m_{i} \epsilon_{jkl} m_{k}
E_{l} .
\label{eq:ah}
\end{eqnarray}
The anomalous Hall conductivity, $\sigma_{\rm AH}$, describes the
charge current due to the anomalous Hall effect, the associated
polarization $\zeta$ expresses the fact that this charge current is
spin polarized.

The anisotropic magnetoresistance\cite{Thomson:1857,McGuire:1975} is
an additional consequence of spin-orbit coupling in ferromagnets.  In
this case, the conductivity of a ferromagnet is different if the
magnetization is along the electric field direction or perpendicular to it.
While not typically considered, the polarization of the conductivity
will change in these two cases.  Another consequence of the anisotropy in
the conductivity occurs when the magnetization is at any angle other
than collinear with or perpendicular to the electric field.
For these other orientations of the magnetization, the charge current has an
additional contribution, which flows in the 
direction of the magnetization.  This current is frequently described
as the planar Hall effect because for a thin film ferromagnet, an
electric field gives rise to a Hall current (perpendicular to the
electric field) when the magnetization is rotated in the plane of the
film.  The charge current direction due to the planar Hall effect is given by
$\mathbf{m}(\mathbf{m}\cdot\mathbf{E})$ and again, the spins flowing
with that current point the ${\bf m}$ direction.  Then, the
anisotropic magnetoresistance gives rise to a spin current 
\begin{eqnarray}
  {\bf Q}
&=& \frac{-\hbar}{2e} \eta \sigma_{\rm AMR} \mathbf{m} \otimes \mathbf{m} (\mathbf{m}\cdot\mathbf{E})
\nonumber\\
Q_{ij} &=& \frac{-\hbar}{2e} \eta \sigma_{\rm AMR} m_{i} m_{j} m_{k} E_{k} .
  \label{eq:amr}
\end{eqnarray}
The conductivity, $\sigma_{\rm AMR}$, describes the difference in the
charge conductivity comparing cases with the magnetic field parallel and
perpendicular to the electric field.  The associated polarization
$\eta$ expresses the fact that this change in the charge current is
spin polarized.  The spins both flow and point along the
magnetization.

The spin currents associated with the anomalous Hall effect and the
anisotropic magnetoresistance can replace those associated with the
spin Hall effect as generators of torques with advantage of being able
to control the orientation of the spins.  Applying an electric field
in the plane of a ferromagnetic layer generates charge and spin
currents flowing perpendicular to it and into adjacent layers.  Thus
in a FM/NM/FM film, where FM and NM refer to ferromagnetic and
non-magnetic layers respectively, an in-plane electric field generates
spin currents flowing perpendicularly to the layers.  These spin
currents exert torques on the magnetizations in both layers.  The
advantage of this approach is the orientation of the flowing spins can
be controlled by varying the directions of the magnetizations.  The
goal of this paper is to evaluate these spin transfer torques and show
how they may be advantageous for some device applications.  We develop
the drift-diffusion equations in Sec.~\ref{sec:deriv} and apply them
to the case in which an electric current flows in the plane of a
FM/NM/FM film.  Details of the derivation are given in the Appendices.
In Sec.~\ref{sec:results}, we illustrate the angular dependence of the
torque as both magnetizations are varied and then show how these
torques can lead to effective magnetization switching and domain wall
motion.  We summarize our results in Sec.~\ref{sec:summary}.


\section{Derivation}
\label{sec:deriv}

In this section we present the drift diffusion equations in ferromagnets,
accounting for the spin-orbit derived contributions to the transport.
Since spin components transverse to the magnetization rapidly precess
and dephase, they can be neglected.  Then, the charge and spin
currents are combinations of the majority and minority currents
carried by spin-$s$ ($s=\uparrow,\downarrow$) electrons.
In the presence of the Anomalous Hall (AH) effect and the anisotropic
magnetoresistance (AMR) effect, the spin current densities are given by 
\begin{eqnarray}
  \mathbf{j}^{\uparrow}
  &=&
  \frac{(1+\beta)}{2}
  \frac{\sigma}{e}
  \bm{\nabla}
  \mu^{\uparrow}
  +
  \frac{(1+\zeta)}{2}
  \frac{\sigma_{\rm AH}}{e}
  \mathbf{m}
  \times
  \bm{\nabla}
  \mu^{\uparrow}
\nonumber\\ &&
  +
  \frac{(1+\eta)}{2}
  \frac{\sigma_{\rm AMR}}{e}
  \mathbf{m}
  \left(
    \mathbf{m}
    \cdot
    \bm{\nabla}
    \mu^{\uparrow}
  \right),
\label{eq:jspinup}\\
  \mathbf{j}^{\downarrow}
  &=&
  \frac{(1-\beta)}{2}
  \frac{\sigma}{e}
  \bm{\nabla}
  \mu^{\downarrow}
  +
  \frac{(1-\zeta)}{2}
  \frac{\sigma_{\rm AH}}{e}
  \mathbf{m}
  \times
  \bm{\nabla}
  \mu^{\downarrow}
\nonumber\\&&  
  +
  \frac{(1-\eta)}{2}
  \frac{\sigma_{\rm AMR}}{e}
  \mathbf{m}
  \left(
    \mathbf{m}
    \cdot
    \bm{\nabla}
    \mu^{\downarrow}
  \right),
\label{eq:jspindn}
\end{eqnarray}
where the (total) electric current density is
$\mathbf{j}=\mathbf{j}^{\uparrow}+\mathbf{j}^{\downarrow}$.   
The longitudinal conductivity and conductivities due to the anomalous Hall
effect and the anisotropic magnetoresistance
effect are denoted as  
$\sigma$, $\sigma_{\rm AH}$, and $\sigma_{\rm AMR}$, respectively, 
and their spin polarizations are denoted as $\beta$, $\zeta$, and
$\eta$ respectively. 
The spin-dependent electro-chemical potentials are denoted as $\mu^{s}$. 
We define electro-chemical potential $\bar{\mu}$ and spin accumulation
$\delta\mu$ as  
\begin{eqnarray}
  \bar{\mu}
  =
  \frac{\mu^{\uparrow}+\mu^{\downarrow}}{2},
&~~~~~~~&
  \delta
  \mu
  =
  \frac{\mu^{\uparrow}-\mu^{\downarrow}}{2}.
\label{eq:mudef}
\end{eqnarray}
We emphasize that the "(longitudinal) spin accumulation" used in
Refs.~\onlinecite{Brataas:2001,Brataas:2006,Tserkovnyak:2005}, which will be used below, is defined
as $\mu^{\uparrow}-\mu^{\downarrow}$, which is twice the magnitude of
$\delta \mu$.  In terms of $\bar{\mu}$ and $\delta\mu$, we find that
\begin{eqnarray}
  \mathbf{j}^{\uparrow}
  +
  \mathbf{j}^{\downarrow}
  &=&
  \frac{\sigma}{e}
  \bm{\nabla}
  \bar{\mu}
  +
  \beta
  \frac{\sigma}{e}
  \bm{\nabla}
  \delta
  \mu
 \nonumber\\ &&
  +
  \frac{\sigma_{\rm AH}}{e}
  \mathbf{m}
  \times
  \bm{\nabla}
  \bar{\mu}
   +
  \zeta
  \frac{\sigma_{\rm AH}}{e}
  \mathbf{m}
  \times
  \bm{\nabla}
  \delta
  \mu
\label{eq:jtot}\\ &&
 +
  \frac{\sigma_{\rm AMR}}{e}
  \mathbf{m}
  \left(
    \mathbf{m}
    \cdot
    \bm{\nabla}
    \bar{\mu} 
  \right)
  +
  \eta
  \frac{\sigma_{\rm AMR}}{e}
  \mathbf{m}
  \left(
    \mathbf{m}
    \cdot
    \bm{\nabla}
    \delta
    \mu
  \right),
\nonumber\\ 
  \mathbf{j}^{\uparrow}
  -
  \mathbf{j}^{\downarrow}
  &=&
  \frac{\sigma}{e}
  \bm{\nabla}
  \delta
  \mu
  +
  \beta
  \frac{\sigma}{e}
  \bm{\nabla}
  \bar{\mu}
 \nonumber\\ &&
  +
  \frac{\sigma_{\rm AH}}{e}
  \mathbf{m}
  \times
  \bm{\nabla}
  \delta
  \mu
  +
  \zeta
  \frac{\sigma_{\rm AH}}{e}
  \mathbf{m}
  \times
  \bm{\nabla}
  \bar{\mu}
 \label{eq:jdiff}\\ &&
  +
  \frac{\sigma_{\rm AMR}}{e}
  \mathbf{m}
  \left(
    \mathbf{m}
    \cdot
    \bm{\nabla}
    \delta
    \mu
  \right)
\ +
  \eta
  \frac{\sigma_{\rm AMR}}{e}
  \mathbf{m}
  \left(
    \mathbf{m}
    \cdot
    \bm{\nabla}
    \bar{\mu}
  \right). \nonumber
\end{eqnarray}
In terms of these current densities, the tensor spin current density is
${\bf Q}=-\frac{\hbar}{2e}{\bf m}\otimes({\bf
  j}^\uparrow-{\bf j}^\downarrow)$.

It is tempting to imagine that all three polarizations, $\beta$,
$\zeta$, and $\eta$ are the same, but there is no reason that they
should be.  The polarization of the longitudinal conductivity, $\beta$
is determined by the spin-dependent densities of states and
particularly the spin-dependent scattering rates.  It is typically
between -1 and 1, with negative values for the rare cases in which the
minority conductivity is higher than the majority.  Values
approach $\pm 1$ for half metals.  Values greater than 1 or less than
-1 would imply that one spin type move backwards.  We are not aware of
any such case.  

The polarizations, $zeta$, that of polarization of the anomalous Hall
effect and $\eta$, that of the anomalous Hall effect are not simply
related to $\beta$.  For example, we can construct several
contradictory arguments for the value of $zeta$.  If we imagine that
the anomalous Hall effect were simply a deflection of all carriers in
one direction and that these carriers then underwent the same
spin-dependent scattering as the longitudinal current, we would guess
that $\zeta\approx\frac{(1+\beta)\sigma_{\rm AH}-{(1-\beta)\sigma_{\rm
      AH}}}{(1+\beta)\sigma_{\rm AH}+{(1-\beta)\sigma_{\rm
      AH}}}=\beta$.  If on the other hand, we imagine that the
anomalous Hall effect originates from the spin Hall effect in which
different spins are deflected in opposite directions and then each
spin is subject to the same spin-dependent scattering, we might
imagine that the majority and minority electrons flow in the opposite
directions but are affected by the same spin dependent scattering as
the conductivity.  The reversed flow for the minority electrons
essentially inverts the polarization
$\zeta\approx\frac{(1+\beta)\sigma_{\rm AH}+{(1-\beta)\sigma_{\rm
      AH}}}{(1+\beta)\sigma_{\rm AH}-{(1-\beta)\sigma_{\rm
      AH}}}=1/\beta$.  In fact, first principles
calculations\cite{privatejulich} of the spin polarization of the
anomalous Hall effect give results that vary widely and do not seem to
agree with any simple model.  Some of this variability can be
understood from first principles calculations\cite{Tanaka:2008} of the
spin Hall effect, which show that the spin Hall conductivity depends
sensitively on the Fermi level.  The spin split-band structure of
ferromagnets can be viewed in a simple approximation as just a shift
in energies of the bands for one spin relative to the other, or
equivalently the two spins see different Fermi energies. In this case,
the minority and majority spins that are deflected in different
directions are deflected by different potentials and will be deflected
in different amounts. Therefore, part of the polarization $\zeta$ of the
anomalous Hall current comes from the energy dependence of the
``underlying spin Hall effect.''  Similarly, $\eta$, the spin
polarization of the anomalous Hall effect, is determined by the change
in the spin-dependent scattering and as such gives no expectation to
its value.

\begin{figure}
\begin{center}
\includegraphics[width=2.75in]{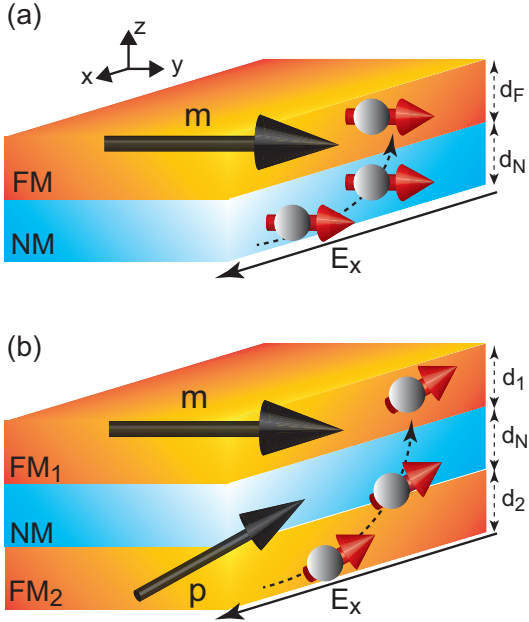}      
\caption{ (color online) (a) Schematic geometry for spin Hall effect
  induced spin transfer torques.  In this geometry, the damping-like
  torque is with respect to the $y$-axis, i.e. ${\bf m}\times(\hat{\bf
    y}\times{\bf m}$ (with a smaller field-like torque). (b) Schematic
  geometry for anomalous Hall effect induced spin transfer torques.
  In this case, the damping-like torque is with respect to the fixed
  layer magnetization direction ${\bf p}$, i.e. ${\bf m}\times({\bf
    p}\times{\bf m})$ (with a smaller field-like torque).  }
\label{fig:geom}
\end{center}
\end{figure}

We are interested in the geometry, illustrated in
Fig.~\ref{fig:geom}(b), in which two ferromagnetic films are separated
from each other by a thin non-magnetic layer that allows the
magnetizations of the two layers to be oriented independently of each
other.  We assume that the interface normals lie in the $z$-direction
and the electric field is applied in the $x$-direction.  We ignore
charge and spin currents that flow in the $y$-direction because they
do not couple to anything.  In general, an electric field in the
$x$-direction would give rise to charge current flow in the z-direction, but
the thin film geometry treated here prevents that.  Except for the
applied electric potential $eE_{x}x$, only the $z$-components of
$\bm{\nabla}\bar{\mu}$ and $\bm{\nabla}\delta\mu$ are non-zero, i.e.,
$\bm{\nabla}(\bar{\mu}/e)=E_{x}\mathbf{e}_{x}+(\partial_{z}\bar{\mu}/e)\mathbf{e}_{z}$
and
$\bm{\nabla}(\delta\mu/e)=(\partial_{z}\delta\mu/e)\mathbf{e}_{z}$.
The electric field adjusts itself so that no electric current flows in
the $z$-direction.


In a particular ferromagnetic layer, we can solve Eqs.~(\ref{eq:jtot})
and (\ref{eq:jdiff}) together with the diffusion
equation\cite{Valet:1993}
\begin{eqnarray}
  \label{eq:spindiff}
  \frac{\partial^{2}}{\partial z^{2}}
  (\mu^{\uparrow} - \mu^{\downarrow})
  &=&
  \frac{\mu^{\uparrow}-\mu^{\downarrow}}{\ell_{\rm sf}^{2}},
\end{eqnarray}
where $\ell_{\rm sf}$ is the spin diffusion length.
In
Appendix~\ref{app:spinacc}, we give the details the derivation of these
solutions.  Here we highlight some of the key steps.
Forcing the charge current in the $z$-direction to be zero dictates that the
spin current in the z-direction have the form 
\begin{equation}
  j_{z}^{\uparrow}
  -
  j_{z}^{\downarrow}
  =
  \tilde{\sigma}_{E}
  E_{x}
  +
  \frac{\tilde{\sigma}_{\delta\mu}}{2e \ell_{\rm sf}}
  \left(
    A
    e^{z/\ell_{\rm sf}}
    -
    B
    e^{-z/\ell_{\rm sf}}
  \right),
\label{eq:sigmas_intro}
\end{equation}
where the constants $A$ and $B$ are to be determined in
Appendix~\ref{app:spinacc}.  The spin current is given
in terms of two
effective conductivities $\tilde{\sigma}_{E}$ and
$\tilde{\sigma}_{\delta\mu}$.  The former essentially gives
the spin current that would result in a bulk material in response to a
field in the $x$-direction in which the
transverse charge current were constrained to be zero.  The latter
gives the spin current in response to a spin accumulation, including
the corrections due to the charge current itself being zero.
The effective conductivities are
\begin{eqnarray}
  \tilde{\sigma}_{E}
  &=&
  \frac{(\beta\sigma + \eta\sigma_{\rm AMR}m_{z}^{2}) (\sigma_{\rm AH}m_{y}-\sigma_{\rm AMR}m_{z}m_{x})}{\sigma + \sigma_{\rm AMR}m_{z}^{2}}
\nonumber\\ &&  
-
  \left(
    \zeta
    \sigma_{\rm AH}
    m_{y}
    -
    \eta
    \sigma_{\rm AMR}
    m_{z}
    m_{x}
  \right),
  \label{eq:sigma_E}
\end{eqnarray}
and
\begin{eqnarray}
  \tilde{\sigma}_{\delta\mu}
  &=&
  \sigma
  +
  \sigma_{\rm AMR}
  m_{z}^{2}
\nonumber\\ &&
  -
  \left(
    \beta
    \sigma
    +
    \eta
    \sigma_{\rm AMR}
    m_{z}^{2}
  \right)
  \left(
    \frac{\beta\sigma+\eta\sigma_{\rm AMR}m_{z}^{2}}{\sigma+\sigma_{\rm AMR}m_{z}^{2}}
  \right).
  \label{eq:sigma_mu}
\end{eqnarray}
While the effective conductivities appear complicated,
$\tilde{\sigma}_{E}$ simplifies considerably in certain limits and
gives simple illustrations of the main results of this paper.  If the
anisotropic magnetoresistance can be neglected, $\tilde{\sigma}_{E}
\rightarrow (\beta-\zeta) m_y \sigma_{\rm AH} $.  Thus, there is a spin
current whenever the magnetization has a component along the
$y$-direction, $Q_{iz}\sim m_i m_y$.  This means that by tilting the
magnetization out-of-plane, it is possible to get an out-of-plane
component the spins flowing into the other layer, something not achievable with the spin
Hall effect in non-magnetic materials.  This feature is illustrated in
Fig~\ref{fig:geom}(b).
The factor of $(\beta-\zeta)$
arises from two contributions, the term proportional to $\zeta$ is
directly from the polarized current accompanying the anomalous Hall
current.  The term proportional to $\beta$ comes from the polarization
of the ``counter-flow'' current that cancels the anomalous Hall
current.

When the anomalous Hall effect can be neglected, $\tilde{\sigma}_{E}
\rightarrow (\eta-\beta) m_x m_z \sigma_{\rm AMR}
\frac{\sigma}{\sigma+\sigma_{\rm AMR}m_z^2}$.  This expression is more
complicated than that for the anomalous Hall effect above because the
anisotropic magnetoresistance affects the conductivity in the
$z$-direction as captured by the last factor in this expression.  As
with the previous case, an out-of plane component of the magnetization
gives an out-of-plane component to the spin current, $Q_{iz}\sim m_i
m_x m_z$.  As with the previous case, the factor of $(\eta-\beta)$
appears from the polarized current due to the planar Hall effect and
the counter-flow current that cancels the charge current of the planar
Hall effect.

Computing the torques on both layers requires finding the
spin accumulation and spin current throughout the structure.
The spin current at the
F${}_{1}$/N interface is given in terms of the spin accumulation at
the F${}_{1}$/N interface and interface conductances.\cite{Brataas:2001,Brataas:2006,Tserkovnyak:2005}
The spin
accumulation  is found by applying 
appropriate boundary conditions to $\bar{\mu}$ and $\delta\mu$ as
described in Appendix~\ref{app:spinacc}.  For a magnetic layer with
interface (1) at $z=0$ and interface (2) at $z=d$, we have
\begin{eqnarray}
  \tilde{\sigma}_{\delta\mu} (\mu^{\uparrow} - \mu^{\downarrow})
  &=&
  \frac{-2e \ell_{\rm sf}}{\sinh(d/\ell_{\rm sf})}
\nonumber\\ &&
\times
  \Bigg[
    \left(
      j_{{\rm s}z}^{(1)}
      -
      \tilde{\sigma}_{E}
      E_{x}
    \right)
    \cosh
    \left(
      \frac{z-d}{\ell_{\rm sf}}
    \right)
\nonumber\\ && ~~~
    -
    \left(
      j_{{\rm s}z}^{(2)}
      -
      \tilde{\sigma}_{E}
      E_{x}
    \right)
    \cosh
    \left(
      \frac{z}{\ell_{\rm sf}}
    \right)
  \Bigg] ,
  \label{eq:solution_spin_accumulation}
\end{eqnarray}
where ${\bf j}_{\rm s}^{(i)}$ is ${\bf j}^\uparrow-{\bf j}^\downarrow$
at the interface of the normal metal with ferromagnet $i$.
The spin current is then
\begin{equation}
\begin{split}
  \mathbf{Q}_{s}^{\rm F_{1} \to N}
  =
  \frac{1}{4\pi}
  &
  \left[
    \frac{(1-\gamma^{2})g}{2}
    \mathbf{m}
    \cdot
    \left(
      \bm{\mu}_{\rm F_{1}}
      -
      \bm{\mu}_{\rm N}
    \right)
    \mathbf{m}
  \right.
\\
  &
  \left.
    -
    g_{\rm r}
    \mathbf{m}
    \times
    \left(
      \bm{\mu}_{\rm N}
      \times
      \mathbf{m}
    \right)
    -
    g_{\rm i}
    \bm{\mu}_{\rm N}
    \times
    \mathbf{m}
  \right] .
  \label{eq:spin_current_F1N_orig}
\end{split}
\end{equation}
Here $g=g^{\uparrow\uparrow}+g^{\downarrow\downarrow}$ and
$\gamma=(g^{\uparrow\uparrow}-g^{\downarrow\downarrow})/g$ are the
dimensionless interface conductance and its spin polarization,
respectively, which relates to the interface resistance $r$ via
$r=[(1/r^{\uparrow\uparrow})+(1/r^{\downarrow\downarrow})]^{-1}=(h/e^{2})S/g$
with $h/e^{2}=25.9$~k$\Omega$.  The cross section area is denoted as
$S$.  The real and imaginary parts of the mixing conductance are
denoted as $g_{\rm r}$ and $g_{\rm i}$, respectively.  Note that the
charge chemical potential does not appear because the fact that the
charge current across the interface is zero allows us to relate the
chemical potential difference to the longitudinal spin chemical
potential difference and eliminate the former from the equation for
the spin current.

The solutions of the spin accumulations in each ferromagnetic layer
and the boundary conditions allow us to write the spin current in each
ferromagnetic layer in terms of the just the spin accumulation in the
non-magnetic layer
\begin{eqnarray}
  \mathbf{Q}_{s}^{\rm F_{1} \to N}
  &=&
  \frac{\hbar g^{*}}{2e g_{\rm sd}^{\prime}}
  \tanh
  \left(
    \frac{d_{1}}{2 \ell_{\rm sf}}
  \right)
  \tilde{\sigma}_{E}
  E_{x}S
  \mathbf{m}
\nonumber\\
  &&-
  \frac{1}{4\pi}
  \left[
    g^{*}
    \big(
      \mathbf{m}
      \cdot
      \bm{\mu}_{\rm N}
    \right)
    \mathbf{m}
\nonumber\\
  &&    +
    g_{\rm r}
    \mathbf{m}
    \times
    \left(
      \bm{\mu}_{\rm N}
      \times
      \mathbf{m}
    \right)
    +
    g_{\rm i}
    \bm{\mu}_{\rm N}
    \times
    \mathbf{m}
  \big],
  \label{eq:spin_current_F1N}
\end{eqnarray}
where $g^{*}$ is defined as 
\begin{equation}
  \frac{1}{g^{*}}
  =
  \frac{2}{(1-\gamma^{2})g}
  +
  \frac{1}{g_{\rm sd}^{\prime}\tanh(d_{1}/\ell_{\rm sf})},
  \label{eq:g_star}
\end{equation}
and
\begin{equation}
  \frac{g_{\rm sd}^{\prime}}{S}
  =
  \frac{h \tilde{\sigma}_{\delta\mu}}{2e^{2}\ell_{\rm sf}}. 
  \label{eq:g_sd}
\end{equation}
Similarly, the spin current at the F${}_{2}$/N interface is given by 
\begin{equation}
\begin{split}
  \mathbf{Q}_{s}^{\rm F_{2} \to N}
  =&
  -\frac{\hbar g^{*}}{2e g_{\rm sd}^{\prime}}
  \tanh
  \left(
    \frac{d_{2}}{2 \ell_{\rm sf}}
  \right)
  \tilde{\sigma}_{E}
  E_{x}S
  \mathbf{p}
\\
  &-
  \frac{1}{4\pi}
  \left[
    g^{*}
    \left(
      \mathbf{p}
      \cdot
      \bm{\mu}_{\rm N}
    \right)
    \mathbf{p}
  \right.
\\
  &
  \left.
    +
    g_{\rm r}
    \mathbf{p}
    \times
    \left(
      \bm{\mu}_{\rm N}
      \times
      \mathbf{p}
    \right)
    +
    g_{\rm i}
    \bm{\mu}_{\rm N}
    \times
    \mathbf{p}
  \right].
  \label{eq:spin_current_F2N}
\end{split}
\end{equation}

In the structure in Fig~\ref{fig:geom}, we separate the two
ferromagnetic layers by a thin non-magnetic layer.  We assume that
this layer effectively breaks the exchange coupling between the two
ferromagnetic layers.  We also assume that it is still thinner than
its mean free path and 
spin diffusion length, so that spin current injected at one interface
transmits unchanged to the other interface.  These assumptions imply
that the spin current and spin accumulation in the spacer layer can be
treated as constant.  This condition means that $\mathbf{Q}_{s}^{\rm
  F_{1} \to N}+\mathbf{Q}_{s}^{\rm F_{2} \to N}=\bm{0}$, from which
$\bm{\mu}_{\rm N}$ can be determined.  Then, the spin torque acting on
$\mathbf{m}$ is obtained from
\begin{equation}
    {\bf T}=\left(\frac{{\rm d} \mathbf{m}}{{\rm d}t}\right)_{\rm st}
    =
    \frac{\gamma_{0}}{\mu_0M_{\rm s}V}
    \mathbf{m}
    \times
    \left(
      \mathbf{Q}_{s}^{\rm F_{1} \to N}
      \times
      \mathbf{m}
    \right),
    \label{eq:spin_torque_F1_def}
\end{equation}
where $\mu_0$ is the magnetic constant and, $\gamma_{0}$, $M_{\rm s}$,
and $V$ are the gyromagnetic ratio, saturation magnetization, and
volume of F${}_{1}$, respectively.  

Further progress requires
taking these solutions for both ferromagnetic layers and solving for
the spin accumulation in the non-magnetic layer.  
In general, the resulting torque can be written in
the form
\begin{eqnarray}
  \label{eq:ofmotion}
{\bf T}
&=&
\frac{\gamma_{0} \hbar E_{x}}{2e \mu_0 M_{\rm s}d_{1}}
\\
&&
\left[
\sigma_{\rm eff}^{\rm d}(\mathbf{m},\mathbf{p}) 
\mathbf{m} \times (\mathbf{p} \times \mathbf{m})
+
\sigma_{\rm eff}^{\rm f}(\mathbf{m},\mathbf{p}) 
\mathbf{p} \times \mathbf{m}
\right] 
\nonumber
\end{eqnarray}
The superscripts on the effective conductivities refer to the
damping-like, $d$, and field-like, $f$, components of the torque.
However, a key point of this paper is that these damping-like and
field-like torques are defined with respect to the orientation of the
magnetization in the other layer, here ${\bf p}$, and not as for the
spin Hall effect, the direction $\hat{\bf E}\times\hat{\bf n}$, where
$\hat{\bf n}$ is the interface normal.  See Fig.~\ref{fig:geom} for
the comparison.  The effective
conductivities depend strongly on the directions of the
magnetizations, ${\bf m}$ and ${\bf p}$.  In particular, they inherit
the strong orientational dependence from $\tilde\sigma_{\rm E}$.  When
the imaginary part of the mixing conductance can be neglected, the
field-like torque vanishes.  The spin torque acting on $\mathbf{p}$ is
obtained in a similar way.  In Appendix~\ref{app:details}, we show how to compute
the torques numerically for the general case and show some analytic
forms for some special cases.  In the next section, we present
numerical results and investigate the consequences of these torques on
switching and domain wall motion.

The derivation in this section is done using the drift-diffusion
approach, as is typically used in the analysis of experiments using
the spin Hall effect to generate spin transfer torques.  This
approximation does not capture the in-plane giant magnetoresistance
effect because in the absence of spin orbit effects, there is no net
spin current flowing from layer to layer.  The simplest calculation to
capture the current-in-plane giant magnetoresistance is based on the
Boltzmann equation.\cite{Camley:1989}  When applied to the spin Hall
effect and resulting torques, this approach\cite{Haney:2013a} yields
quantitative but not qualitative differences in comparison with the
drift diffusion approach.  We expect the same to be true for the
present calculations.  It is also the case that the in-plane giant
magnetoresistance, in the absence of spin-orbit coupling, does not
lead to a spin transfer torque even though spin flow from each layer
to the other.


\section{Results}
\label{sec:results}


\subsection{Angular dependence of torques}

\begin{figure*}
\begin{center}
\includegraphics[width=5.5in]{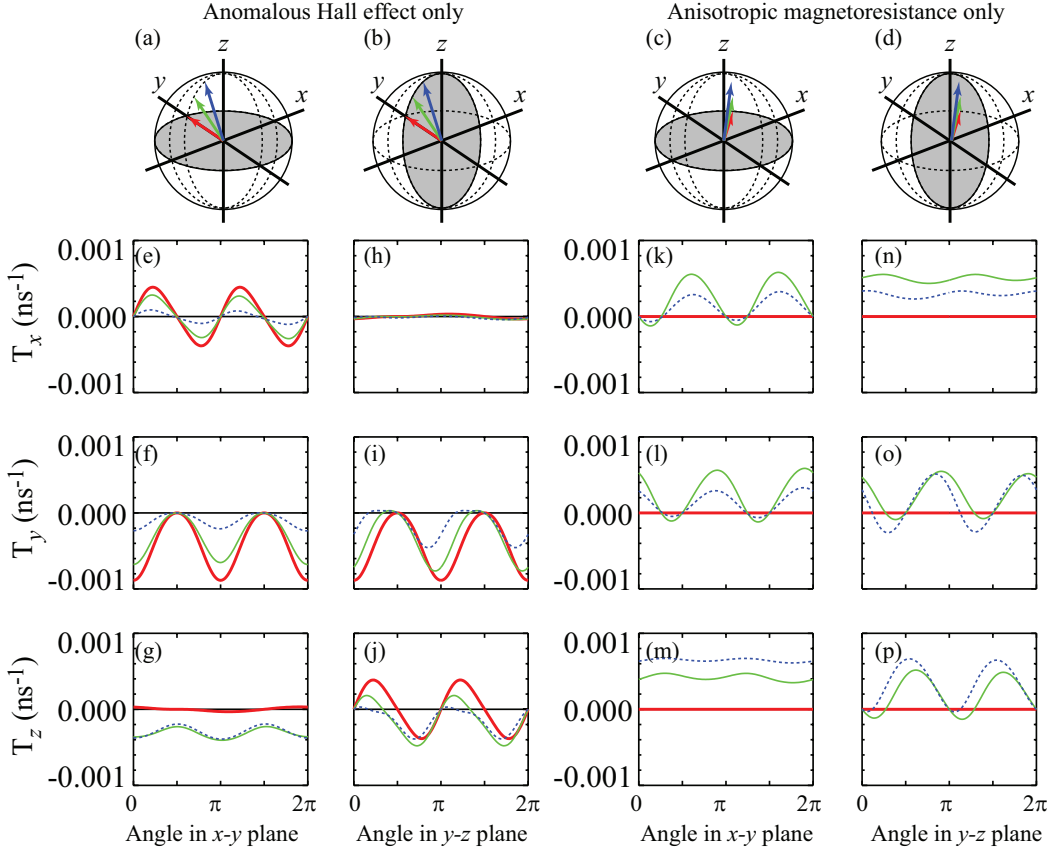}      
\caption{ (color online) Angular dependence of spin transfer torques.
  Panels (a) through (d) show the directions of the magnetizations in
  each column.  The gray projected circles indicate the plane of
  rotation of the free layer magnetization, ${\bf m}$.  For panels
  (a) and (c) the plane of rotation is the $x-y$ plane, starting at
  $\hat{\bf x}$ and for panels (b) and (d) it is the $y-z$ plane
  starting at $\hat{\bf z}$ .  The arrows in panels (a-d) indicate the
  three directions of the fixed layer magnetization, ${\bf p}$ for
  the panels in each column (online, the colors correspond to the
  colors of the curves in the panels below).  These are at $\theta=
  90^\circ,\ 60^\circ,\ {\rm and}\ 30^\circ$ for all four panels and
  $\phi= 90^\circ$ for (a) and (b) and $\phi= 45^\circ$ for (c) and
  (d).  Panels (e-j) give the torques for the anomalous Hall effect
  with the anisotropic magnetoresistance set to zero and panels (k-p)
  the other way around.  In each of the panels (e-p) the heavy (red)
  lines give the torque for $\theta=90^\circ$, light (green) lines for
  $\theta=60^\circ$ and dashed (blue) for $\theta=30^\circ$.  Rows
  (e,h,k,n), (f,i,l,o), and (g,j,m,p) give the $x$, $y$, and $z$
  components of the torque respectively.  For all calculations, the
  current density is $10^{11}$~A/m$^2$.  }
\label{fig:angle}
\end{center}
\end{figure*}

While the full solution of the torque for a general model is quite
complicated, it can be qualitatively understood much more simply.
Using the parameters in Table~\ref{tab:matparm}, we compute the torque
for a variety of magnetization directions for two 5 nm thick NiFe
layers and plot them in
Fig.~\ref{fig:angle}.  For simplicity, we consider two cases,
$\sigma_{\rm AMR}=0$ and $\sigma_{\rm AH}=0$, so we can show the
effect of each separately.  In the limit that both are much less than
$\sigma$, the two contributions should add.

\begin{table}
\begin{tabular}{lrlrlrll}
\hline\noalign{\smallskip}
 & NiFe && CoFeB && FePt && units \\
\hline\noalign{\smallskip}
$\rho$                 & 122&$^{\rm a}$   & 300&$^{\rm b}$ & 390&$^{\rm c}$ &  $\Omega$nm \\
$\beta$                & 0.7&$^{\rm a}$   & 0.56&$^{\rm b}$ & 0.40&$^{\rm d}$ & \\ 
$r$                    & 0.5&$^{\rm a}$   & 0.5&$^{\rm b}$ & 0.5 && k$\Omega$nm${}^{2}$ \\ 
$\gamma$               & 0.7&$^{\rm a}$  & 0.83&$^{\rm b}$ & 0.83 && \\ 
$g_{\rm r}/S$            & 10.0&$^{\rm e}$  & 10.0 && 10.0 && nm${}^{-2}$ \\ 
$g_{\rm i}/S$            & 1.0   && 0.0 && 0.0 && nm${}^{-2}$ \\ 
$\ell_{\rm sf}$          & 5.5&$^{\rm a}$   & 4.5&$^{\rm f}$ & 5.0&$^{\rm d}$ & nm \\ 
$\sigma_{\rm AH}/\sigma$ & 0.001&$^{\rm g}$ & 0.0 && 0.015&$^{\rm c}$ & \\
$\sigma_{\rm AMR}/\sigma$ & 0.06&$^{\rm h}$ & 0.0 && 0.0147&$^{\rm i}$ & \\
$\zeta$                & 5    && 0 && 1.5 && \\ 
$\eta$                 & 0.9  && 0 && -0.1 &&  \\ 
$M_{\rm s}$              & 0.86&$^{\rm j}$ & 0.456&$^{\rm k}$& && MA/m \\ 
$H_{\rm K}$              & 0.0&  & 0.569&$^{\rm k}$& && MA/m \\
$\gamma_{0}$            & 0.23206  && 0.23206 && && Mm/(A s) \\ 
$\alpha$               & 0.01&$^{\rm j}$  & 0.01 && && \\
\noalign{\smallskip}\hline
\end{tabular}
\caption{Default material parameters.  Parameters are chosen to
  approximate Ni$_{80}$Fe$_{20}$ (Permalloy), CoFeB and FePt, but
  some values are not well known.  In particular, $\eta$ and $\zeta$
  are unknown to our knowledge and so we have chosen representative
  values.  Values for parameters are taken from
(a)~Ref.~\onlinecite{Bass:1999}, 
(b)~Ref.~\onlinecite{Oshima:2002}
(c)~Ref.~\onlinecite{Moritz:2008}, 
(d)~Ref.~\onlinecite{Seki:2008},
(e)~Ref.~\onlinecite{Bauer:2003}
(f)~Ref.~\onlinecite{Ahn:2008},
(g)~Ref.~\onlinecite{Zhang:2013},
(h)~Ref.~\onlinecite{Rijks:1997},
(i)~Ref.~\onlinecite{Christides:1994}, 
(j)~Ref.~\onlinecite{Tannenwald:1957},
(k)~Ref.~\onlinecite{Zhang:2012}, 
(l)~Ref.~\onlinecite{Zhou:2008} where indicated and estimated where
not indicated.
} 
\label{tab:matparm}
\end{table}

Consider first the case in which there is only the anomalous Hall
effect.  We have assumed that the imaginary part of the mixing
conductance is much less than the real part, so any field-like torque
that is present is also much smaller than the damping-like
contribution.  The discussion in Sec.~\ref{sec:intro} that
$\tilde\sigma_{\rm E}~(\beta-\zeta) p_y \sigma_{\rm AH}$ for the spin
current due to the fixed layer with its magnetization in the ${\bf
  p}$ direction, gives guidance for the approximate angular dependence
of the torque.  Since the spins in the spin current point in the
${\bf p}$ direction, the damping-like torque varies like $p_y
{\bf m} \times ( {\bf p} \times {\bf m})$.  When the
magnetization is along the $y$-axis, the torque has the same angular
dependence as the spin Hall effect as seen in the heavy (red) curves
of Fig.~\ref{fig:angle}(e-j). 
That is, a damping-like torque with respect to the $y$-axis.  In this
case, the out-of-plane torque, $T_z$, (heavy red curve in
Fig.~\ref{fig:angle}(g)) is essentially zero when the 
magnetization is rotated in plane.

As the fixed layer magnetization is rotated out of plane (light (green)
and dashed (blue) curves in Fig.~\ref{fig:angle}(e-j)), the torque
remains damping-like, $p_y
{\bf m} \times ( {\bf p} \times {\bf m})$, but it develops an
out-pf-plane component, $T_z$, even when the magnetization is rotated in
plane, (light (green)
and dashed (blue) curves in Fig.~\ref{fig:angle}(g).  This breaks the
symmetry between ${\bf m}=\pm\hat{\bf z}$, 
making it possible to reliably switch the magnetization, as discussed
in the next section.  However, as the polarizer magnetization is
rotated toward the pole, the total size of the torque goes to zero
because $p_y$ 
goes to zero when $p_z\rightarrow \pm 1$.

When the anomalous Hall effect is absent and the anisotropic
magnetoresistance is present (Fig.~\ref{fig:angle}(k-p)), the angular
dependence is slightly more 
complicated.  Recall from Sec.~\ref{sec:intro} that
$\tilde\sigma_{\rm E} \rightarrow (\eta-\beta) p_x p_z \sigma_{\rm AMR}
\frac{\sigma}{\sigma+\sigma_{\rm AMR}p_z^2}$ when the anomalous Hall
effect is absent.  If $\sigma_{\rm AMR}/\sigma \ll 1$, the last factor
can be neglected.  In that case, the damping-like torque varies
like $p_x p_z {\bf m} \times ( {\bf p} \times {\bf m})$.
The spin current flows along the magnetization
direction, so unless $p_z \ne 0$ there is no spin current flow into
the free layer.  Thus, the torque is zero when
the fixed layer magnetization is in-plane (heavy (red) curves in
Fig.~\ref{fig:angle}(k-p)).  Otherwise, it has roughly 
a damping-like form with respect to the fixed layer magnetization.
For the values of parameters we have assumed, there are deviations
from the simple ${\bf m} \times ( {\bf p} \times {\bf m})$
behavior expected when the spin-orbit effects are weak.


\subsection{Magnetic Switching}

One advantage of spin-orbit effects in ferromagnets, as compared to
the spin Hall effect, is that the control over the direction of the
incident spin current allows for the excitation of magnetization
dynamics that cannot be excited by the spin Hall effect.  An example
of such dynamics is a switching of a perpendicularly magnetized free
layer in the absence of an external field.  In this section, we
analytically compute the critical current for switching a
perpendicular magnetization in F$_1$ due to the anomalous
Hall effect and anisotropic magnetoresistance effect in F${}_{2}$.  
We verify the behavior by direct numerical
simulation of the Landau-Lifshitz-Gilbert (LLG) equation. 

For illustrative purposes, we simplify the generally complex 
dependence on relative angle of the magnetizations seen in
Eq.~(\ref{eq:torque_F1}) by treating a special case.  
We assume that F${}_{1}$ has neither the anomalous Hall effect
nor the anisotropic magnetoresistance, 
i.e., $\sigma_{\rm AH(F_{1})}=\sigma_{\rm AMR(F_{1})}=0$, 
whereas F${}_{2}$ has both. 
The magnetization of F${}_{1}$, $\mathbf{m}$, can move freely, 
whereas that of F${}_{2}$, $\mathbf{p}$, points to an
arbitrary fixed direction. 
The values of the parameters are taken from 
CoFeB free (F${}_{1}$) layer and 
FePt pinned (F${}_{2}$) layer, 
and summarized in Table \ref{tab:matparm}.

The LLG equation for the magnetization in 
F${}_{1}$, with the spin torque, Eq. (\ref{eq:ofmotion}), is
\begin{equation}
\begin{split}
  \frac{{\rm d}\mathbf{m}}{{\rm d}t}
  =&
  -\gamma_{0}
  \mathbf{m}
  \times
  \mathbf{H}
  +
  \alpha
  \mathbf{m}
  \times
  \frac{{\rm d}\mathbf{m}}{{\rm d}t}
\\
  &
  +
  \frac{\gamma_{0}\hbar }{2e\mu_{0}M_{\rm s}d_{1}}
  E_{x}
  \sigma_{\rm eff}^{\rm d}
  \mathbf{m}
  \times
  \left(
    \mathbf{p}
    \times
    \mathbf{m}
  \right),
  \label{eq:LLG_macrospin_AHE}
\end{split}
\end{equation}
where $\alpha$ is the Gilbert damping constant, 
and $\sigma_{\rm eff}^{\rm d}$ is given by (see also Appendix~\ref{app:details}) 
\begin{equation}
  \sigma_{\rm eff}^{\rm d}
  =
  \frac{\tanh[d_{2}/(2\ell_{\rm sf}^{\rm F_{2}})] g_{\rm r(F_{1})} g_{\rm F_{2}}^{*}(\mathbf{p}) \tilde{\sigma}_{E({\rm F}_{2})}(\mathbf{p})}
    {g_{\rm sd(F_{2})}^{\prime}(\mathbf{p}) (g_{\rm r(F_{1})} + g_{\rm F_{2}}^{*}(\mathbf{p})) [1 - \lambda_{1} \lambda_{2}(\mathbf{p}) (\mathbf{m}\cdot\mathbf{p})^{2}]}. 
\end{equation}
We introduce the parameter $\lambda_{k}$ ($k=1,2$), 
which characterizes the dependence of the spin torque strength 
on the relative angle of the magnetizations, 
\begin{equation}
  \lambda_{k} 
  = 
  \frac{g_{{\rm r}({\rm F}_{k})} - g_{{\rm F}_{k}}^{*}}{g_{{\rm r}({\rm F}_{k^{\prime}})} + g_{{\rm F}_{k}}^{*}},
  \label{eq:lambda_def}
\end{equation}
where $(k,k^{\prime})=(1,2)$ or $(2,1)$. 
We emphasize that $g_{\rm sd(F_{2})}^{\prime}(\mathbf{p})$, 
$g_{\rm F_{2}}^{*}(\mathbf{p})$, $\lambda_{2}(\mathbf{p})$, 
and $\tilde{\sigma}_{E({\rm F}_{2})}(\mathbf{p})$ 
depend on the direction of $\mathbf{p}$, 
according to their definition, Eqs. (\ref{eq:sigma_E}), (\ref{eq:sigma_mu}), (\ref{eq:g_star}), (\ref{eq:g_sd}), and (\ref{eq:lambda_def}). 
On the other hand, $\lambda_{1}$ is independent of $\mathbf{m}$ 
because the F${}_{1}$ layer does not show the anomalous Hall effect nor anisotropic magnetoresistance effect. 

We assume that F${}_{1}$ is a perpendicular magnet with an anisotropy
field given by $\mathbf{H}=(0,0,(H_{\rm K}-M_{\rm s})m_{z})$, 
where $H_{\rm K}$ is the perpendicular anisotropy field. 
In the absence of an electric field $E_{x}$, 
the free layer magnetization is stable along the perpendicular
axis.  We assume that it starts along the $z$-axis, 
i.e., $\mathbf{m}=\hat{\bf z}$. 
In the presence of the spin torque, 
the magnetization is destabilized, 
and starts to precess around the $z$-axis. 
Assuming that $m_{z}\simeq 1$ and $|m_{x}|,|m_{y}| \ll 1$, 
we can linearize the LLG equation (see Appendix~\ref{app:perpswitch})
and determine the critical current
\begin{equation}
\begin{split}
  j_{\rm crit}
  =&
  -\frac{2\alpha e \mu_{0} M_{\rm s} d_{1} (H_{\rm K}-M_{\rm s})}{\hbar\tanh[d_{2}/(2 \ell_{\rm sf}^{\rm F_{2}})]}
\\
  &\times
  \frac{(1-\lambda_{1}\lambda_{2} p_{z}^{2})^{2} g_{\rm sd(F_{2})}^{\prime} (g_{\rm r(F_{1})} + g_{\rm F_{2}}^{*}) \sigma_{\rm F_{2}}}
    {(1-\lambda_{1}\lambda_{2}) p_{z} g_{\rm F_{2}}^{*} g_{\rm r(F_{1})} \tilde{\sigma}_{E({\rm F}_{2})}}. 
  \label{eq:critical_current} 
\end{split}
\end{equation}

\begin{figure}
\begin{center}
\includegraphics[width=2.75in]{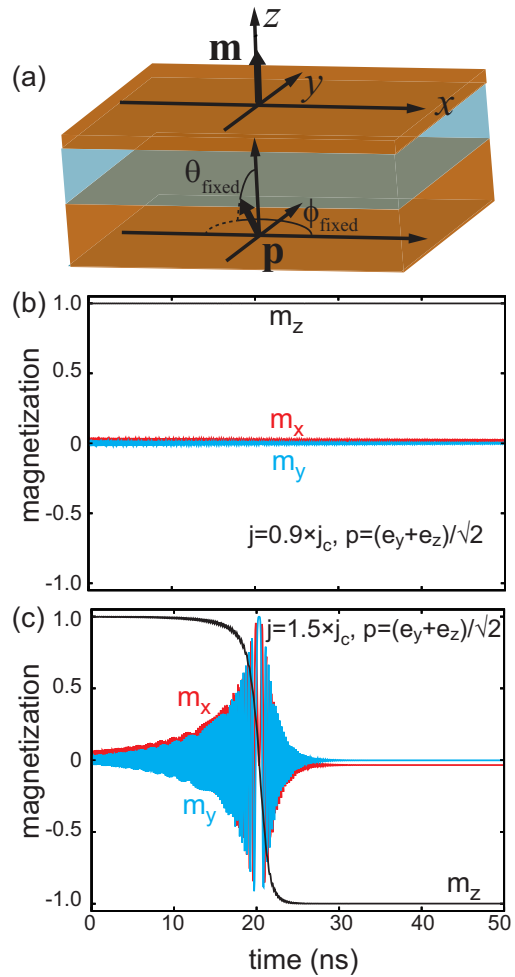}      
\caption{ (color online) Magnetization dynamics due to the anomalous
  Hall effect.  Panel (a) shows the geometry.  The trajectories
  obtained by numerically solving the LLG equation
  (\ref{eq:LLG_macrospin_AHE}) are shown in (b) for $j=0.9 \times
  j_{\rm c}$ and (c) for $j=1.5 \times j_{\rm c}$.  }
\label{fig:macrospin_LLG}
\end{center}
\end{figure}

Using Eq.~(\ref{eq:critical_current}), we can estimate the
critical current for field-free switching of perpendicular layers. 
As an example, let us assume that F${}_{2}$ has 
the anomalous Hall effect only, i.e., $\sigma_{\rm AH(F_{2})} \neq 0$ and $\sigma_{\rm AMR(F_{2})}=0$. 
In this case, $\tilde{\sigma}_{E({\rm F}_{2})}$ is $(\beta_{\rm
  F_{2}}-\zeta_{\rm F_{2}})p_{y}\sigma_{\rm AH(F_{2})}$ and
Eq.~(\ref{eq:critical_current}) can be simplified to
Eq.~(\ref{eq:critical_current_AHE}).  
We choose the pinned layer magnetization
to be $\mathbf{p}=(0,1/\sqrt{2},1/\sqrt{2})$ and take the
parameter values given in
Table~\ref{tab:matparm}.
For 10 nm of  FePt, which can be fixed in a partially out of
plane configuration, as a polarizer and 1 nm of CoFeB, with
perpendicular anisotropy, as a free layer, 
we find a critical current of 
$1.0 \times 10^{12}$~A/m${}^{2}$ from Eq.~(\ref{eq:critical_current}).
In Fig.~\ref{fig:macrospin_LLG}, we show the magnetization dynamics
obtained by numerically solving the LLG equation
(\ref{eq:LLG_macrospin_AHE}) for the electric current densities of (a) $j=0.9 \times j_{\rm c}$ 
and (b) $j=1.5 \times j_{\rm c}$, respectively.  
The magnetization stays near the initial direction in (a), 
whereas it switches the direction to $\mathbf{m}=-\hat{\mathbf{z}}$, 
showing the validity of Eq.~(\ref{eq:critical_current}). 


\begin{figure}
\begin{center}
\includegraphics[width=2.75in]{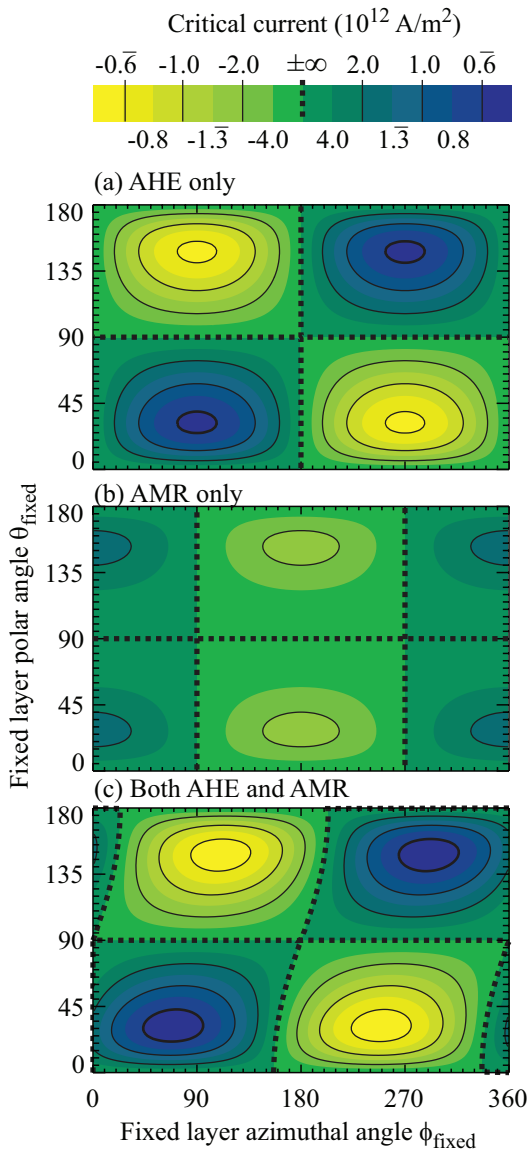}      
\caption{ (color online) Critical currents for a CoFeB free layer and
  FePt fixed layer as a function of the fixed layer magnetization
  direction.  The contours are chosen uniformly in the inverse critical
  current, the contour where the critical currents diverge is labeled
  $\pm\infty$.  In panel (a), we assume that the polarizer has
  anomalous Hall effect (AHE) but no anisotropic magnetoresistance
  (AMR).  In panel (b), we assume it has the AMR but no AHE, and in
  panel (c) we assume it has both.  Dark (blue) regions indicate
  regions with low critical for one direction of current flow and
  light (yellow) regions indicate low critical currents for the other
  direction.  At the equator ($\theta_{\rm fixed}=90^\circ$), the
  critical currents diverge for all three cases, however, for the case
  with only AMR [panel (b)], the sign does not change as $\theta_{\rm
    fixed}$ is varied near that point, but for the other two cases it
  does.}
\label{fig:critical_currents}
\end{center}
\end{figure}

Figure \ref{fig:critical_currents} shows the switching current as a
function of the orientation of the fixed layer magnetization
$\mathbf{p}=(\sin(\theta_{\rm fixed})\cos(\phi_{\rm
  fixed}),\sin(\theta_{\rm fixed})\sin(\phi_{\rm
  fixed}),\cos(\theta_{\rm fixed}))$ from
Eq.~(\ref{eq:critical_current}), and verified by numerical simulation
of the LLG equation.  The three panels show switching due to the
anomalous Hall effect and anisotropic magnetoresistance separately and
combined.  For the parameters chosen here, given in
Table~\ref{tab:matparm}, the anomalous Hall effect is more efficient.
The figure shows that the most efficient switching occurs when the
polarizer magnetization is close to perpendicular ($\theta_{\rm
  fixed}\approx 0^\circ$).  The efficiency is
determined by a competition between two effects.  One effect is the
efficiency of the spins at destabilizing the magnetization toward
reversal.  Spins injected perpendicular to the stable magnetization
direction exert the greatest torque, but since they enhance precession
only over half a period and suppress it over the other, they do not
destabilize the magnetization.  Electrons with moments antiparallel to
the magnetization exert no torque, but when the magnetization
fluctuates, they exert a torque that destabilizes the magnetization
over the whole precession period.  When the critical current is large
enough, they overcome the damping and any fluctuations get magnified,
leading to reversal.  The counterbalancing effect is that when the
pinned layer magnetization is collinear with the magnetization, it is
also collinear with the film normal and the injected spin current goes
to zero.  So, the most efficient switching occurs with the pinned
layer magnetization close to normal but not all the way there,
maximizing the total perpendicular component of the injected spins.
Switching due to the anomalous Hall 
effect and that due to anisotropic magnetoresistance depend
differently on the azimuthal angle so for some orientations of the
fixed layer magnetization, they compete, but for others they cooperate
to reduce the critical current.

The critical current is minimized at an optimal direction of
$\mathbf{p}$.  Because of complex dependences of $\tilde{\sigma}_{E}$
and $\tilde{\sigma}_{\delta\mu}$ on the magnetization direction, as
shown in Eqs. (\ref{eq:sigma_E}) and (\ref{eq:sigma_mu}), it is
difficult to derive a formula of this optimal direction.  However, for
the F${}_{2}$ with the anomalous Hall effect only, we can derive the
analytical formula of the optimum direction of $\mathbf{p}$; see
Appendix~\ref{app:optimum}.  The result, for this set of parameters is
$\theta_{\rm fixed}=31.6^{\circ}$, $\phi_{\rm fixed}=90^\circ$.


We can compare these results with 
the magnetization switching assisted by the spin Hall effect. 
In the spin Hall effect, spin current polarized along the
$\hat{\mathbf{y}}$ direction  
is injected to the free layer. 
This situation is similar to a special case of switching by spin-orbit
effects in ferromagnets 
in which the pinned layer magnetization is in the $\hat{\bf y}$
direction.  It is useful to consider a generalized situation with the
fixed layer magnetization
in the $yz$-plane, $\phi_{\rm fixed}= 90^\circ$ with no anisotropic
magnetoresistance.  
Then, $\tilde\sigma_{\rm E}$ simplifies and
Eq.~(\ref{eq:critical_current}) has the factor $p_{y}p_{z} $ in the
denominator as seen in Eq.~(\ref{eq:critical_current_AHE}).
This factor implies that $j_{\rm c}^{\rm AH}$ diverges when 
$\mathbf{p}$ points to the $z$-direction ($p_{y}=0$ and $p_{z}=1$)
because the anomalous Hall effect does not induce spin current along
the $z$-direction  
when $p_{y}=0$. 
The critical current also diverges when $\mathbf{p}$ points to the
$y$-direction ($p_{y}=1$ and $p_{z}=0$) because the spin-transfer
torque never overcomes the damping torque as needed to enhance
precession.  This is the equivalent of switching by the spin Hall
effect.  While the spin-transfer torque can excite
magnetization dynamics, when the fixed layer magnetization is along
$\hat{\bf y}$ it does not overcome the damping and does not cause
precession to become unstable.

It is possible to excite dynamics in perpendicularly magnetized
samples with the spin Hall effect (or the anomalous Hall effect with
${\bf p}=\hat{\bf y}$) as shown by Lee \textit{et
  al.}\cite{Lee:2013}.  In fact, they demonstrate that it is possible
to switch the magnetization.  However, the switching they observe is
not due to the spin transfer torque overcoming the damping, but rather
is due to a large amplitude excitation due to the rapid onset of the
current and hence torque.  However, since nothing in the system breaks
the symmetry between up and down, such switching is extremely
sensitive to pulse duration and current amplitude.  Lee \textit{et
  al.}\cite{Lee:2013} demonstrate such sensitivity in Fig.~1(b) of
their paper.  They derive an analytic form, Eq.~(5), for the critical
current that is independent of the damping parameter.  This
independence indicates that the switching mechanism is precessional,
rather than due to overcoming damping.  
To switch the magnetization direction without such sensitivity, 
an in-plane magnetic field slightly tilted to the $z$-direction has been used 
experimentally \cite{Liu:2012b}. 
The switching mechanism due to
the anomalous Hall effect with a fixed layer with an out-of-plane
component to the magnetization has the advantage of being largely
independent of the current density or pulse duration for currents
above the critical current.  
Another advantage is that the external field is unnecessary to switch
the magnetization.  
It can also be significantly lower when
the damping parameter is small, as is desirable in many magnetic
devices.


\subsection{Domain wall motion}

\begin{figure}
\begin{center}
	\includegraphics[width=2.75in]{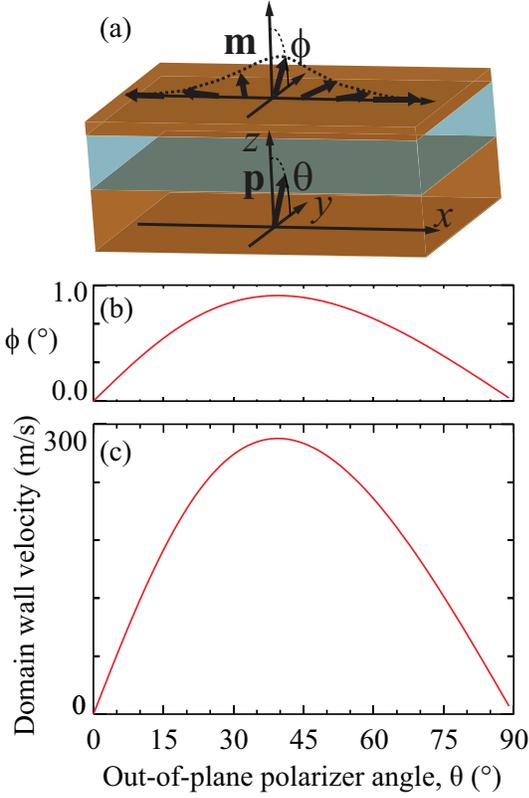}
        \caption{(color online) (a) Schematic of the spin valve with a
          transverse in-plane wall and a fixed uniform polarizer. (b)
          Out-of-plane tilt angle of domain wall and (c) Domain wall
          velocity as a function of the out-of-plane angle $\theta$ of
          the polarizer.  Calculations are done for 10~nm Py for a
          polarizer layer, 1~nm Py for a free layer, and a charge current
          density of $2\times 10^{11}$~A/m$^2$.}
	\label{fig:DW1}
\end{center}
\end{figure}

The spin-orbit torques generated by ferromagnets can also be useful to
displace in-plane magnetic domain walls, which we illustrate through
two simple examples.  We first consider the spin-valve illustrated in
Fig.~\ref{fig:DW1}(a), with an in-plane domain wall in the free layer
F1 and a uniform polarizer $\mathbf{p} = (0, p_y,p_z)$ in the fixed
layer F2. Due to the spin orbit effects in F2, a torque is generated
on F1 that has the form : $\mathbf{T} =  \tau_{\rm
  so}(\mathbf{m},\mathbf{p}) ~
\mathbf{m}\times(\mathbf{m}\times\mathbf{p})$. To study the effect of
this torque we consider a 1D model\cite{Schryer:1974} of a transverse wall
profile with a domain wall width $\Delta$.  The magnetization in the
free layer, with the domain wall, is subject to a spin current from a
fixed layer below.  This spin current will cause a small tilting of
the magnetization away from the long axis in all of the domains and
will cause motion of the domain wall.  We neglect the small tilting of
the domains to get the following equations for the domain wall dynamics: 
\begin{eqnarray}
\label{1Dmodel-STT}
\dot{\phi}+\frac{\alpha}{\Delta}\dot{q}&=&  \tau_{\rm so} p_z \cos\phi -  \tau_{\rm so} p_y \sin\phi\\ 
\frac{\dot{q}}{\Delta}-\alpha\dot{\phi}&=& \gamma_0 H_k \sin\phi \cos\phi 
\end{eqnarray}
Here $q$ is the domain wall position, $\phi$ the out-of-plane tilt
angle and $H_k$ the shape anisotropy.  At equilibrium in the absence
of spin torques, $\phi$ is equal to zero and the domain wall lies in
plane.  

In the regime below Walker breakdown, the wall moves with a
constant tilt angle and 
a steady velocity.  Assuming the tilt is small, $\sin\phi \ll 1$,
\begin{eqnarray}\label{velocity-AH-singleDW}
\phi &=& \frac{\tau_{\rm so} p_z}{\alpha \gamma_0 H_k + \tau_{\rm so} p_y}
\nonumber\\
\dot{q}_{\rm AH} &=& \frac{\Delta}{\alpha} \tau_{\rm so}
p_z \frac{\alpha \gamma_0 H_k}{\alpha \gamma_0 H_k + \tau_{\rm so} p_y} 
\end{eqnarray}
Since $\alpha \gamma_0 H_k \gg \tau_{\rm so}$ for typical values of
the current density, the out-of-plane tilt is indeed small.
The domain wall moves
steadily only if the generated spin torque has a component along
the $z$-direction. This is not the case of the torque generated by
pure spin Hall effect in a non-magnetic heavy metal, in which case the
domain wall does not move.\cite{Khvalkovskiy:2013} On the other hand,
the spin-orbit torques 
generated by a ferromagnet can have components along both the $z$ and
$y$ directions
when the polarizer is tilted out-of-plane. 
If, as we did in the last
section, we consider the case of
the torque generated by just the anomalous Hall effect in F2, then
\begin{eqnarray}
  \label{eq:efficency}
\tau_{\rm AH} &=& \frac{\gamma_0 \hbar}{2e\mu_0 M_{\rm s}} 
\frac{\tanh[d_2/(2 \ell_{\rm sf})]}{d_1}
\frac{g^{*} g_{\rm r}} {g_{\rm
    sd}^{\prime}(g_{\rm r}+g^{*})} 
\nonumber\\ &&
\frac{1}{1-\lambda^{2}(\mathbf{m}\cdot\mathbf{p})^{2}}
(\beta-\zeta)\sigma_{\rm AH} E_{x} p_{y}  
\end{eqnarray}
This behavior is shown in Fig.~\ref{fig:DW1}, in which we treat the
motion for the case with the anomalous Hall effect and anisotropic
magnetoresistance in both layers.  However, since we assume the
magnetization lies in the $y-z$ plane, the anisotropic
magnetoresistance plays a negligible role.  Fig.~\ref{fig:DW1} shows a
relatively large domain wall velocity for a modest charge current density of
$2\times 10^{11}$~A/m$^2$ and a very small out of plane tilt of less than a
degree. 

In the proposed spin-valve system,
the current flowing in the ferromagnet $F_1$ through the domain wall
will also give rise to the more familiar (intralayer) adiabatic and
non-adiabatic 
spin-transfer torques on the domain wall, these
can enhance or oppose the effect of the spin-orbit torques. In
comparison, the domain wall velocity induced by these intralayer torques is :
\begin{equation}
\label{velocity-na-singleDW}
\dot{q}_{\rm na} = \frac{1}{\alpha} \frac{\gamma_0 \hbar}{2e\mu_0 M_{\rm s}}
P \beta_{\rm na} \sigma E_{x} 
\end{equation}
where $P\approx\beta$ is the current polarization and
$\beta_{\rm na}$ the 
proportionality factor between the non-adiabatic and adiabatic
torques. The ratio of the velocities is
\begin{equation}
\label{comp-na-AH}
\frac{\dot{q}_{\rm AH}}{\dot{q}_{\rm na}}  \approx  \frac{\Delta}{d_1}
\frac{p_{y} p_{z}(\beta-\zeta)\sigma_{\rm AH} }{P \beta_{\rm na} \sigma} 
F ,
\end{equation}
where $F$ is a series of factors (see Eq.~(\ref{eq:efficency}) of
order one.
In a typical material as NiFe, both the anomalous spin hall angle and
the non-adiabatic parameter $\beta_{\rm na}$ are close to 1~$\%$
\cite{Miao:2013}. However, the domain wall will be mainly driven by
the 
anomalous Hall torque because the wall width is typically much bigger
than the layer thickness $\Delta/d> 10$ for most
systems.\cite{Metaxas:2013} 

\begin{figure}
\begin{center}
	\includegraphics[width=2.75in]{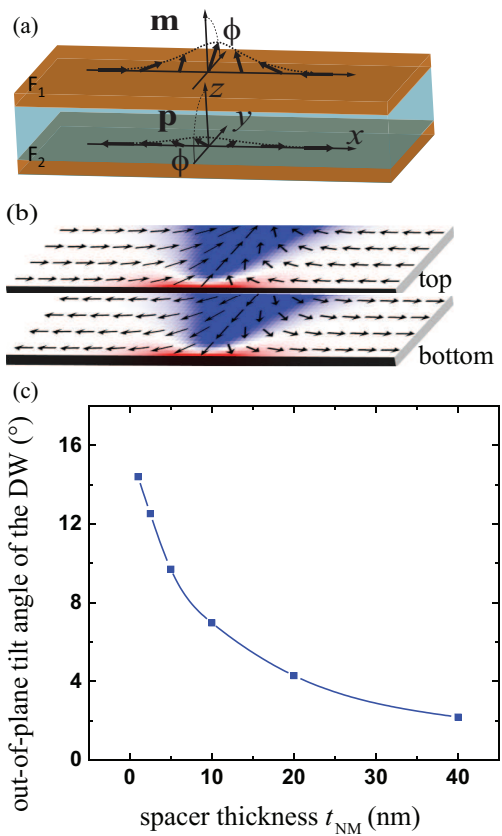}
        \caption{(color on-line) (a) Schematic of the spin valve in
          anti-parallel 
          configuration with two domain walls, one in each layer. (b)
          micromagnetic simulations of the coupled domain wall
          system showing in blue the positive z component of the
          magnetization. (c) Maximum out-of-plane tilt angle
          $\phi$ of the magnetization in the domain wall as a function
          of the spacer thickness.}
	\label{fig:DW2}
\end{center}
\end{figure}

The other system we consider is the coupled domain
wall system shown in Fig.~\ref{fig:DW2}(a). In the case of a fixed
polarizer $F_2$ and a free layer $F_1$, $F_2$ can exert a torque on
$F_1$. But if $F_2$ is no longer fixed, $F_1$ can also induce a torque
on $F_2$. If the magnetic configuration is well chosen, these
reciprocal torques can add and enhance magnetization dynamics of the
coupled system. This is the case for the double domain wall system
with anti-parallel configuration shown in Fig.~\ref{fig:DW2}.

If both magnetic layers are unpinned, the domain walls in each
layer are strongly coupled.  Domain walls in wires with opposite
in-plane magnetizations tilt out of plane significantly due to the
dipolar interaction between them, as shown in Fig.\ref{fig:DW2}.
In equilibrium, one domain wall has the out-of-plane tilt angle
$\phi_0$, and the other $\pi-\phi_0$ so that the out-of-plane
component is in the 
same direction and the in-plane directions are opposite.
This configuration is illustrated in the micromagnetic simulations in
Fig.~\ref{fig:DW2}(b) where blue shows the
out-of-plane component of the magnetization.\cite{micromagdetails} As
the spacer thickness $t_{NM}$  
decreases, the dipolar fields on each domain wall increase, and the
maximum out-of-plane tilt angle increases as shown in
Fig.~\ref{fig:DW2}(b), reaching values close to 15$^\circ$ for spacer
thicknesses typical of synthetic antiferromagnets.

In this configuration, the domain wall in $F_2$ polarizes the domain wall in $F_1$
(and reciprocally), and we can replace $p_y$ and $p_z$ respectively by
$-cos\phi$ and $sin\phi$ in Eq.~(\ref{1Dmodel-STT}). For small angle
deviations from the equilibrium configuration, this immediately leads
to
\begin{equation}\label{velocity-AH-doubleDW}
\dot{q}_{\rm AH} =  \frac{\Delta}{\alpha} \tau_{\rm so} \sin(2 \phi_0)
\end{equation}
Due to the particular symmetry of the anomalous Hall effect
torques, the domain wall in $F_2$ acquires the same velocity: the motion of the
coupled domain wall system is self-sustained. For small spacer
thicknesses, the tilt angle is large, and velocities comparable to the
single wall system with a uniform tilted fixed polarizer can be
reached.


\section{Summary}
\label{sec:summary}

In this paper we develop a drift-diffusion approach to treat
transport effects of spin-orbit coupling in ferromagnets.  These
include the anomalous Hall effect and the anisotropic
magnetoresistance.  In addition to the transverse charge currents that
arise due to these effects, there are concomitant spin currents.
These spin currents flow perpendicularly
to the electric field, and so can be injected into layers
perpendicular to the electrical current flow.  When these other layers
are ferromagnets with magnetizations that are not aligned with the
original layer, they create spin transfer torques.  Unlike the related
spin Hall effect in non-magnetic materials, the ferromagnetic
spin-orbit effects allow some control of
the orientation of the injected spins.  This control arises because
the flowing spins in a ferromagnet are collinear with the
magnetization.  Changing the orientation of the magnetization changes
the direction of the spins injected into other layers.

We compute the torques due to current flow for two ferromagnet
layers separated by a thin non-magnetic layer.  The control of the
direction of the injected spins makes it is possible to switch
perpendicularly magnetized layers more easily because of the
possibility of an out-of-plane component of the torque.  We also
show that such torques make it possible to switch in-plane
magnetized layers via propagation of transverse/vortex walls and can
efficiently induce dynamics in coupled magnetic systems, e.g. coupled
transverse domain walls.

\begin{acknowledgments}
The authors thank Robert McMichael for useful discussions.  JG
acknowledges funding from the European Research Council Grant No.~259068. 
\end{acknowledgments}

\begin{widetext}

\appendix


\section{Solution of electro-chemical potential and spin accumulation}
\label{app:spinacc}

The $x$ and $z$-components of Eq.~(\ref{eq:jtot}) 
 are explicitly given terms of $\bar{\mu}$ and $\delta \mu$ by 
\begin{equation}
  j_{x}
  =
  \sigma
  E_{x}
  +
  \frac{\sigma_{\rm AH}}{e}
  (\partial_{z}\bar{\mu})
  m_{y}
  +
  \sigma_{\rm AMR}
  \left[
    E_{x}
    m_{x}
    +
    \frac{1}{e}
    (\partial_{z}\bar{\mu})
    m_{z}
  \right]
  m_{x}
+
  \zeta
  \frac{\sigma_{\rm AH}}{e}
  (\partial_{z}\delta\mu)
  m_{y}
  +
  \eta
  \frac{\sigma_{\rm AMR}}{e}
  (\partial_{z}\delta\mu)
  m_{z}
  m_{x},
\end{equation}
\begin{equation}
  j_{z}
  =
  \frac{\sigma}{e}
  \partial_{z}
  \bar{\mu}
  -
  \sigma_{\rm AH}
  E_{x}
  m_{y}
  +
  \sigma_{\rm AMR}
  \left[
    E_{x}
    m_{x}
    +
    \frac{1}{e}
    (\partial_{z}\bar{\mu})
    m_{z}
  \right]
  m_{z}
+
  \beta
  \frac{\sigma}{e}
  \partial_{z}
  \delta
  \mu
  +
  \eta
  \frac{\sigma_{\rm AMR}}{e}
  (\partial_{z}\delta\mu)
  m_{z}^{2}.
\end{equation}
The continuity equation for electric current in steady state,
$\bm{\nabla}\cdot\mathbf{j}=\partial_{z}j_{z}=0$, requires $(\sigma +
\sigma_{\rm AMR} m_{z}^{2}) \bar{\mu} + (\beta \sigma + \eta
\sigma_{\rm AMR} m_{z}^{2}) \delta \mu = Cz + D + F(x)$, where $C$ and
$D$ are the integral constants whereas $F(x) \propto eE_{x}x$.  The
condition $j_{z}=0$ implies $C= e (\sigma_{\rm AH}m_{z} - \sigma_{\rm
  AMR} m_{z} m_{x}) E_{x}$, whereas the other integral constant $D$
corresponds to a shift of the chemical potential, $\mu_{\rm shift}$.
Then, the electro-chemical potential is
\begin{eqnarray}
  \bar{\mu}
  =
  \mu_{\rm shift}
  +
  e E_{x}x
  +
  \left(
    \frac{\sigma_{\rm AH}m_{y}-\sigma_{\rm AMR}m_{z}m_{x}}{\sigma+\sigma_{\rm AMR}m_{z}^{2}}
  \right)
  e E_{x} z 
  -
  \frac{1}{2}
  \left(
    \frac{\beta\sigma + \eta\sigma_{\rm AMR}m_{z}^{2}}{\sigma + \sigma_{\rm AMR}m_{z}^{2}}
  \right)
  \left( \mu^\uparrow - \mu^\downarrow
  \right). 
\label{eq:electrochem}
\end{eqnarray}
We assume that the spin accumulation obeys the diffusion equation,
Eq.~(\ref{eq:spindiff}).  The solution can be expressed as
$\mu^{\uparrow}-\mu^{\downarrow}=A e^{z/\ell_{\rm sf}} + B
e^{-z/\ell_{\rm sf}}$.  Two integral constants, $A$ and $B$, are
determined as follows.  Using Eq.~(\ref{eq:electrochem}),
the $z$-component of Eq.~(\ref{eq:jdiff}) is
Eq.~(\ref{eq:sigmas_intro}) and
the spin current is 
$-[\hbar/(2e)](j_{z}^{\uparrow}-j_{z}^{\downarrow})$.  When the
ferromagnet lies in the region $0 \le z \le d$, and the spin current densities
at $z=0$ and $d$ are given by $j_{sz}^{(1)}$ and $j_{sz}^{(2)}$,
respectively, the integral constants, $A$ and $B$ are determined as
\begin{equation}
  \frac{\tilde{\sigma}_{\delta\mu}}{2e\ell_{\rm sf}}
  A
  =
  \frac{1}{2 \sinh(d/\ell_{\rm sf})}
  \left[
    j_{{\rm s}z}^{(1)}
    e^{-d/\ell_{\rm sf}}
    -
    j_{{\rm s}z}^{(2)}
    -
    \tilde{\sigma}_{E}
    \left(
      1
      -
      e^{-d/\ell_{\rm sf}}
    \right)
    E_{x}
  \right],
\end{equation}
\begin{equation}
  \frac{\tilde{\sigma}_{\delta\mu}}{2e\ell_{\rm sf}}
  B
  =
  \frac{1}{2 \sinh(d/\ell_{\rm sf})}
  \left[
    j_{{\rm s}z}^{(1)}
    e^{d/\ell_{\rm sf}}
    -
    j_{{\rm s}z}^{(2)}
    +
    \tilde{\sigma}_{E}
    \left(
      e^{d/\ell_{\rm sf}}
      -
      1
    \right)
    E_{x}
  \right].
\end{equation}
These give Eq.~(\ref{eq:solution_spin_accumulation}).  In the geometry
shown in Fig. \ref{fig:geom}, the spin current at the F/N interface is
$-\mathbf{m}\cdot\mathbf{Q}_{s}^{\rm F_{1} \to N}$ or
$\mathbf{p}\cdot\mathbf{Q}_{s}^{\rm F_{2} \to N}$, and it is zero
at the outer boundaries.  Using these boundary conditions,
Eq.~(\ref{eq:spin_current_F1N_orig}) can be rewritten as
Eq.~(\ref{eq:spin_current_F1N}). Note that
$(j_{z}^{\uparrow}-j_{z}^{\downarrow})$ satisfies
\begin{equation}
  e
  \frac{\partial (j_{z}^{\uparrow}-j_{z}^{\downarrow})}{\partial z}
  =
  \frac{(\sigma+\sigma_{\rm AMR}m_{z}^{2})^{2}-(\beta\sigma+\eta\sigma_{\rm AMR}m_{z}^{2})^{2}}
    {(\sigma+\sigma_{\rm AMR}m_{z}^{2})}
  \frac{(\mu^{\uparrow}-\mu^{\downarrow})}{2\ell_{\rm sf}^{2}},
\end{equation}
which becomes
$[e/(1-\beta^{2})\sigma]\partial(j_{z}^{\uparrow}-j_{z}^{\downarrow})/\partial
z=\delta\mu/\ell_{\rm sf}^{2}$ in the absence of the AMR effect,
reproducing the diffusion equation in Ref.~\onlinecite{Valet:1993}.


\section{Details of the Calculation}
\label{app:details}

The spin current is calculated from Eqs. (\ref{eq:spin_current_F1N}) and (\ref{eq:spin_current_F2N}) 
by assuming the conservation of the spin current inside the N layer, 
i.e., $\mathbf{Q}_{s}^{\rm F_{1} \to N} + \mathbf{Q}_{s}^{\rm F_{2} \to N}=\bm{0}$. 
This condition leads to the following equations 
to determine the components of $\bm{\mu}_{\rm N}$; 
\begin{equation}
  \mathsf{M}
  \begin{pmatrix}
    \mu_{x} \\
    \mu_{y} \\
    \mu_{z}
  \end{pmatrix}
  =
  -\frac{4\pi \hbar g_{\rm F_{2}}^{*}(\mathbf{p})}{2e g_{\rm sd(F_{2})}^{\prime}(\mathbf{p})}
  \tanh
  \left(
    \frac{d_{2}}{2 \ell_{\rm sf}^{\rm F_{2}}}
  \right)
  \tilde{\sigma}_{E({\rm F_{2}})}(\mathbf{p})
  E_{x}S
  \begin{pmatrix}
    p_{x} \\
    p_{y} \\
    p_{z}
  \end{pmatrix}
  +
  \frac{4\pi \hbar g_{\rm F_{1}}^{*}(\mathbf{m})}{2e g_{\rm sd(F_{1})}^{\prime}(\mathbf{m})}
  \tanh
  \left(
    \frac{d_{1}}{2 \ell_{\rm sf}^{\rm F_{1}}}
  \right)
  \tilde{\sigma}_{E({\rm F_{1}})}(\mathbf{m})
  E_{x}S
  \begin{pmatrix}
    m_{x} \\
    m_{y} \\
    m_{z}
  \end{pmatrix}. 
  \label{eq:equation_for_spin_accumulation}
\end{equation}
Here, the components of the $3\times 3$ matrix $\mathsf{M}$ are given by 
\begin{align}
&
  \mathsf{M}_{1,1}
  =
  g_{\rm F_{1}}^{*} 
  m_{x}^{2} 
  +
  g_{\rm r(F_{1})} 
  \left(
    1 
    - 
    m_{x}^{2} 
  \right) 
  +
  g_{\rm F_{2}}^{*} 
  p_{x}^{2} 
  +
  g_{\rm r(F_{2})} 
  \left(
    1 
    - 
    p_{x}^{2} 
  \right), 
\\
&
  \mathsf{M}_{1,2}
  =
  \left(
    g_{\rm F_{1}}^{*} 
    -
    g_{\rm r(F_{1})} 
  \right)
  m_{x} 
  m_{y} 
  +
  g_{\rm i(F_{1})} 
  m_{z} 
  +
  \left(
    g_{\rm F_{2}}^{*} 
    -
    g_{\rm r(F_{2})} 
  \right)
  p_{x} 
  p_{y} 
  +
  g_{\rm i(F_{2})} 
  p_{z}, 
\\
&
  \mathsf{M}_{1,3}
  =
  \left(
    g_{\rm F_{1}}^{*} 
    -
    g_{\rm r(F_{1})} 
  \right)
  m_{z} 
  m_{x} 
  -
  g_{\rm i(F_{1})} 
  m_{y} 
  +
  \left(
    g_{\rm F_{2}}^{*} 
    -
    g_{\rm r(F_{2})} 
  \right)
  p_{z} 
  p_{x} 
  -
  g_{\rm i(F_{2})} 
  p_{y}, 
\\
&
  \mathsf{M}_{2,1}
  =
  \left(
    g_{\rm F_{1}}^{*} 
    -
    g_{\rm r(F_{1})} 
  \right)
  m_{x} 
  m_{y} 
  -
  g_{\rm i(F_{1})} 
  m_{z} 
  +
  \left(
    g_{\rm F_{2}}^{*} 
    -
    g_{\rm r(F_{2})} 
  \right) 
  p_{x} 
  p_{y}
  -
  g_{\rm i(F_{2})} 
  p_{z}, 
\\
&
  \mathsf{M}_{2,2}
  =
  g_{\rm F_{1}}^{*} 
  m_{y}^{2} 
  +
  g_{\rm r(F_{1})} 
  \left(
    1 
    - 
    m_{y}^{2} 
  \right) 
  +
  g_{\rm F_{2}}^{*} 
  p_{y}^{2} 
  +
  g_{\rm r(F_{2})} 
  \left( 
    1 
    - 
    p_{y}^{2} 
  \right), 
\\
&
  \mathsf{M}_{2,3}
  =
  \left(
    g_{\rm F_{1}}^{*} 
    -
    g_{\rm r(F_{1})} 
  \right) 
  m_{y} 
  m_{z} 
  +
  g_{\rm i(F_{1})} 
  m_{x} 
  +
  \left(
    g_{\rm F_{2}}^{*} 
    -
    g_{\rm r(F_{2})} 
  \right) 
  p_{y} 
  p_{z} 
  +
  g_{\rm i(F_{2})} 
  p_{x}, 
\\
&
  \mathsf{M}_{3,1}
  =
  \left(
    g_{\rm F_{1}}^{*} 
    -
    g_{\rm r(F_{1})} 
  \right) 
  m_{z} 
  m_{x} 
  +
  g_{\rm i(F_{1})} 
  m_{y} 
  +
  \left(
    g_{\rm F_{2}}^{*} 
    -
    g_{\rm r(F_{2})} 
  \right) 
  p_{z} 
  p_{x} 
  +
  g_{\rm i(F_{2})} 
  p_{y}, 
\\
&
  \mathsf{M}_{3,2}
  =
  \left(
    g_{\rm F_{1}}^{*} 
    -
    g_{\rm r(F_{1})} 
  \right) 
  m_{y} 
  m_{z} 
  -
  g_{\rm i(F_{1})} 
  m_{x} 
  +
  \left(
    g_{\rm F_{2}}^{*} 
    -
    g_{\rm r(F_{2})} 
  \right)
  p_{y} 
  p_{z} 
  -
  g_{\rm i(F_{2})} 
  p_{x}, 
\\
&
  \mathsf{M}_{3,3}
  =
  g_{\rm F_{1}}^{*} 
  m_{z}^{2}
  +
  g_{\rm r(F_{1})} 
  \left(
    1 
    - 
    m_{z}^{2} 
  \right) 
  +
  g_{\rm F_{2}}^{*} 
  p_{z}^{2} 
  +
  g_{\rm r(F_{2})} 
  \left( 
    1 
    - 
    p_{z}^{2}
  \right). 
\end{align}
The solution of $\bm{\mu}_{\rm N}=(\mu_{x},\mu_{y},\mu_{z})$ can be
obtained by calculating the inverse of $\mathsf{M}$.  In
Eq.~(\ref{eq:equation_for_spin_accumulation}), we added
"$(\mathbf{p})$" and "$(\mathbf{m})$" after $g^{*}$, $g_{\rm
  sd}^{\prime}$, and $\tilde{\sigma}_{E}$ to emphasize that these
quantities depend explicitly on the magnetization direction through
Eqs. (\ref{eq:sigma_E}), (\ref{eq:sigma_mu}), (\ref{eq:g_star}), and
(\ref{eq:g_sd}).  From $\bm{\mu}$ we evaluate the spin currents,
Eqs. (\ref{eq:spin_current_F1N}) and (\ref{eq:spin_current_F2N}).  The
LLG equations for $\mathbf{m}$ and $\mathbf{p}$
are, respectively, given by
\begin{align}
&
  \frac{{\rm d} \mathbf{m}}{{\rm d} t}
  =
  -\gamma_{0}
  \mathbf{m}
  \times
  \mathbf{H}
  +
  \frac{\gamma_{0}}{\mu_{0}M_{\rm s}V}
  \mathbf{m}
  \times
  \left(
    \mathbf{Q}_{s}^{\rm F_{1} \to N}
    \times
    \mathbf{m}
  \right)
  +
  \alpha
  \mathbf{m}
  \times
  \frac{{\rm d} \mathbf{m}}{{\rm d} t},
\\
&
  \frac{{\rm d} \mathbf{p}}{{\rm d} t}
  =
  -\gamma_{0}
  \mathbf{p}
  \times
  \mathbf{H}
  +
  \frac{\gamma_{0}}{\mu_{0} M_{\rm s}V}
  \mathbf{p}
  \times
  \left(
    \mathbf{Q}_{s}^{\rm F_{2} \to N}
    \times
    \mathbf{p}
  \right)
  +
  \alpha
  \mathbf{p}
  \times
  \frac{{\rm d} \mathbf{p}}{{\rm d} t},
\end{align}
where $\gamma_{0}$ and $\alpha$ are the gyromagnetic ratio and Gilbert
damping constant, respectively.  The volume is $V$.


\subsection{Special cases for the spin torque}
\label{app:special}

Although it is possible to solve 
Eq.~(\ref{eq:equation_for_spin_accumulation}) analytically for an arbitrary magnetization alignment, 
the solution looks complicated. 
However, relatively simple analytical formulas can be obtained in some special cases.  
In this section, we discuss such cases.  
Note that Eq.~(\ref{eq:equation_for_spin_accumulation}) 
comes from the conservation law for spin current inside the normal metal
layer, $\mathbf{Q}_{s}^{\rm F_{1} \to N} + \mathbf{Q}_{s}^{\rm F_{2}
  \to N}=\bm{0}$, which can be written as
\begin{equation}
\begin{split}
  &
  g_{\rm F_{1}}^{*}
  (\mathbf{m}\cdot\bm{\mu}_{\rm N})
  \mathbf{m}
  +
  g_{\rm r(F_{1})}
  \mathbf{m}
  \times
  \left(
    \bm{\mu}_{\rm N}
    \times
    \mathbf{m}
  \right)
  +
  g_{\rm i(F_{1})}
  \bm{\mu}_{\rm N}
  \times
  \mathbf{m}
\\
  &
  +
  g_{\rm F_{2}}^{*}
  (\mathbf{p}\cdot\bm{\mu}_{\rm N})
  \mathbf{p}
  +
  g_{\rm r(F_{2})}
  \mathbf{p}
  \times
  \left(
    \bm{\mu}_{\rm N}
    \times
    \mathbf{p}
  \right)
  +
  g_{\rm i(F_{2})}
  \bm{\mu}_{\rm N}
  \times
  \mathbf{p}
\\
  &=
  s_{1}
  \mathbf{m}
  -
  s_{2}
  \mathbf{p},
  \label{eq:equation_for_spin_accumulation_sub}
\end{split}
\end{equation}
where $s_{k}=[4\pi \hbar g_{{\rm F}_{k}}^{*}/(2e g_{{\rm sd}({\rm F}_{k})}^{\prime})] \tanh[d_{k}/(2 \ell_{\rm sf}^{{\rm F}_{k}})] \tilde{\sigma}_{E({\rm F}_{k})} E_{x} S$ ($k=1,2$): 
see Eq.~(\ref{eq:equation_for_spin_accumulation}). 
We expand $\bm{\mu}_{\rm N}$ as 
\begin{equation}
  \bm{\mu}_{\rm N}
  =
  a_{m}
  \mathbf{m}
  +
  b_{m}
  \mathbf{m}
  \times
  \mathbf{p}
  +
  c_{m}
  \mathbf{m}
  \times
  \left(
    \mathbf{p}
    \times
    \mathbf{m}
  \right). 
\end{equation}
Substituting this expression into
Eq.~(\ref{eq:equation_for_spin_accumulation_sub}), and using the
simplification $g_{\rm i}=0$, the coefficients $a_{m}$, $b_{m}$, and
$c_{m}$ are 
\begin{align}
&
  a_{m}
  =
  \frac{ [(g_{\rm r(F_{1})} + g_{\rm F_{2}}^{*}) + (g_{\rm r(F_{2})} - g_{\rm F_{2}}^{*}) (\mathbf{m}\cdot\mathbf{p})^{2} ] s_{1} - (g_{\rm r(F_{1})} + g_{\rm r(F_{2})}) \mathbf{m}\cdot\mathbf{p} s_{2}}
    {(g_{\rm r(F_{1})} + g_{\rm F_{2}}^{*}) (g_{\rm r(F_{2})} + g_{\rm F_{1}}^{*}) - (g_{\rm r(F_{1})} - g_{\rm F_{1}}^{*}) (g_{\rm r(F_{2})} - g_{\rm F_{2}}^{*}) (\mathbf{m}\cdot\mathbf{p})^{2}},
\\
&
  b_{m}
  =
  0,
\\
  &
  c_{m}
  =
  \frac{(g_{\rm r(F_{2})}-g_{\rm F_{2}}^{*}) \mathbf{m}\cdot\mathbf{p} s_{1} - (g_{\rm r(F_{2})}+g_{\rm F_{1}}^{*}) s_{2}}
    {(g_{\rm r(F_{1})} + g_{\rm F_{2}}^{*}) (g_{\rm r(F_{2})} + g_{\rm F_{1}}^{*}) - (g_{\rm r(F_{1})} - g_{\rm F_{1}}^{*}) (g_{\rm r(F_{2})} - g_{\rm F_{2}}^{*}) (\mathbf{m}\cdot\mathbf{p})^{2}}.
\end{align}
The spin torque acting on the magnetization of
the F${}_{1}$ layer, $\mathbf{m}$, is $[\gamma_{0}/(\mu_{0} M_{\rm s}V)] \mathbf{m}
\times (\mathbf{Q}_{s}^{\rm F_{1} \to N} \times \mathbf{m})
=-[\gamma_{0}g_{\rm r}/(4\pi \mu_{0} M_{\rm s}V)] \mathbf{m} \times (\bm{\mu}_{\rm N}
\times \mathbf{m})$.  Then, the coefficient $c_{m}$
and its direction $\mathbf{m}\times(\mathbf{p}\times\mathbf{m})$
gives the spin torque.  The explicit form of the spin torque acting on
$\mathbf{m}$ is
\begin{equation}
\begin{split}
  \frac{{\rm d} \mathbf{m}}{{\rm d} t}
  =&
  \frac{\gamma_{0} \hbar E_{x}}{2e \mu_{0} M_{1} d_{1}} 
  g_{\rm r(F_{1})}
  \frac{\mathbf{m} \times (\mathbf{p} \times \mathbf{m})}{1-\lambda_{1}(\mathbf{m}) \lambda_{2}(\mathbf{p}) (\mathbf{m}\cdot\mathbf{p})^{2}},
\\
  &\times
  \left\{
    \frac{g_{\rm F_{2}}^{*}(\mathbf{p}) \tanh[d_{2}/(2 \ell_{\rm sf}^{\rm F_{2}})] \tilde{\sigma}_{E({\rm F}_{2})}(\mathbf{p})}
      {g_{\rm sd(F_{2})}^{\prime}(\mathbf{p}) [g_{\rm r(F_{1})} + g_{\rm F_{2}}^{*}(\mathbf{p})]}
    -
    \lambda_{2}(\mathbf{p})
    \frac{g_{\rm F_{1}}^{*}(\mathbf{m}) \tanh[d_{2}/(2\ell_{\rm sf}^{\rm F_{1}})] \tilde{\sigma}_{E({\rm F}_{1})}(\mathbf{m})}
      {g_{\rm sd(F_{1})}^{\prime}(\mathbf{m}) [g_{\rm r(F_{1})} + g_{\rm F_{2}}^{*}(\mathbf{p})]}
    \mathbf{m}
    \cdot
    \mathbf{p}
  \right\}
  \label{eq:torque_F1}
\end{split}
\end{equation}
where $\lambda_{k}$ is defined by Eq. (\ref{eq:lambda_def}). 
Note that the conductance
$g^{*}$ and $g_{\rm sd}^{\prime}$, and therefore $\lambda$, depend on not only 
the material parameters but also the magnetization direction 
when the anisotropic magnetoresistance effect is finite; see Eqs. (\ref{eq:sigma_mu}), (\ref{eq:g_star}), 
and (\ref{eq:g_sd}).  
Also, $\tilde{\sigma}_{E}$ depends on the magnetization direction, as shown in Eq. (\ref{eq:sigma_E}). 
Therefore, we add "$(\mathbf{m})$" or "$(\mathbf{p})$" after $g_{{\rm F}_{k}}^{*}$,
$g_{{\rm sd}({\rm F}_{k})}$, $\tilde{\sigma}_{E({\rm F}_{k})}$, and $\lambda_{k}$ 
to emphasize the fact that these
depend on the magnetization direction, $\mathbf{m}$ or $\mathbf{p}$.
Similarly, the spin torque acting on the magnetization of the F${}_{2}$ layer is given by 
\begin{equation}
\begin{split}
  \frac{{\rm d} \mathbf{p}}{{\rm d} t}
  =&
  -\frac{\gamma_{0} \hbar E_{x}}{2e \mu_{0} M_{2} d_{2}} 
  g_{\rm r(F_{2})}
  \frac{\mathbf{p} \times (\mathbf{m} \times \mathbf{p})}{1-\lambda_{1}(\mathbf{m}) \lambda_{2}(\mathbf{p}) (\mathbf{m}\cdot\mathbf{p})^{2}},
\\
  &\times
  \left\{
    \frac{g_{\rm F_{1}}^{*}(\mathbf{m}) \tanh[d_{1}/(2 \ell_{\rm sf}^{\rm F_{1}})] \tilde{\sigma}_{E({\rm F}_{1})}(\mathbf{m})}
      {g_{\rm sd(F_{1})}^{\prime}(\mathbf{m}) [g_{\rm r(F_{2})} + g_{\rm F_{1}}^{*}(\mathbf{m})]}
    -
    \lambda_{1}(\mathbf{m})
    \frac{g_{\rm F_{2}}^{*}(\mathbf{p}) \tanh[d_{2}/(2\ell_{\rm sf}^{\rm F_{2}})] \tilde{\sigma}_{E({\rm F}_{2})}(\mathbf{p})}
      {g_{\rm sd(F_{2})}^{\prime}(\mathbf{p}) [g_{\rm r(F_{2})} + g_{\rm F_{1}}^{*}(\mathbf{m})]}
    \mathbf{m}
    \cdot
    \mathbf{p}
  \right\}
  \label{eq:torque_F2}
\end{split}
\end{equation}
These formulas can be simplified in the absence of the anisotropic
magnetoresistance effect, which we show in the following sections.



\subsubsection{When $\sigma_{\rm AMR}=0$ and only the F${}_{2}$ has an
  anomalous Hall effect}

In the absence of the anisotropic
magnetoresistance effect, 
i.e., $\sigma_{\rm AMR}=0$, 
$g^{*}$, $g_{\rm sd}^{\prime}$, and $\lambda$ become independent from the magnetization directions. 
In this section, we also assume that the material parameters are
identical between two ferromagnets, for simplicity.  In this case,
many of the derived parameters become independent of the layer and
we suppress those indices.

Since $\tilde{\sigma}_{E}$ of the F${}_{1}$ layer is zero
and that of the F${}_{2}$ layer is $\tilde{\sigma}_{E({\rm
    F}_{2})}=(\beta-\zeta)\sigma_{\rm AH}p_{y}$.  The conductance
$g_{\rm sd}$, Eq.~(\ref{eq:g_sd}), and $g^{*}$,
Eq.~(\ref{eq:g_star}), are independent of the magnetization direction
because $\tilde{\sigma}_{\delta\mu}=(1-\beta^{2})\sigma$ is
independent of the magnetization direction.  Then, from
Eq.~(\ref{eq:torque_F1}), the spin torque acting on
$\mathbf{m}$ is
\begin{equation}
\begin{split}
  \frac{{\rm d}\mathbf{m}}{{\rm d} t}
  &=
  \frac{\gamma_{0} \hbar}{2eM_{\rm s}d}
  \frac{g^{*} g_{\rm r} (\beta-\zeta) \tanh[d/(2 \ell_{\rm sf})] \sigma_{\rm AH} E_{x}}
    {g_{\rm sd}^{\prime}(g_{\rm r}+g^{*})}
  p_{y}
  \frac{\mathbf{m}\times(\mathbf{p}\times\mathbf{m})}{1-\lambda^{2}(\mathbf{m}\cdot\mathbf{p})^{2}}. 
  \label{eq:spin_torque_F1_AHE_asymmetric}
\end{split}
\end{equation}
Similarly, the spin torque acting on the F${}_{2}$ layer,
$\mathbf{p}$, is obtained from Eq.~(\ref{eq:torque_F2})
as
\begin{equation}
\begin{split}
  \frac{{\rm d}\mathbf{p}}{{\rm d} t}
  &=
  \frac{\gamma_{0} \hbar}{2e \mu_{0} M_{\rm s}d}
  \frac{g^{*} g_{\rm r} (\beta-\zeta) \tanh[d/(2 \ell_{\rm sf})] \sigma_{\rm AH} E_{x}}
    {g_{\rm sd}^{\prime}(g_{\rm r}+g^{*})}
  p_{y}
  \lambda
  \mathbf{m}
  \cdot
  \mathbf{p}
  \frac{\mathbf{p}\times(\mathbf{m}\times\mathbf{p})}{1-\lambda^{2}(\mathbf{m}\cdot\mathbf{p})^{2}}.
  \label{eq:spin_torque_F2_AHE_asymmetric}
\end{split}
\end{equation}


\subsubsection{When $\sigma_{\rm AMR}=0$ and both the F${}_{1}$ and
  F${}_{2}$ layers show the anomalous Hall effect}

In this case, $\tilde{\sigma}_{E}$ of the F${}_{1}$ and F${}_{2}$
layers are given by $(\beta-\zeta)\sigma m_{y}$ and
$(\beta-\zeta)\sigma p_{y}$, respectively.  The spin torques acting on
$\mathbf{m}$ and $\mathbf{p}$ are obtained from
Eqs. (\ref{eq:torque_F1}) and
(\ref{eq:torque_F2}) as
\begin{equation}
  \frac{{\rm d}\mathbf{m}}{{\rm d} t}
  =
  \frac{\gamma_{0} \hbar}{2e\mu_{0} M_{\rm s}d}
  \frac{g^{*} g_{\rm r} (\beta-\zeta) \tanh[d/(2 \ell_{\rm sf})] \sigma_{\rm AH} E_{x}}
    {g_{\rm sd}^{\prime}(g_{\rm r}+g^{*})}
  \left[
    \frac{p_{y} - m_{y} \lambda \mathbf{m}\cdot\mathbf{p}}{1 - \lambda^{2} (\mathbf{m}\cdot\mathbf{p})^{2}} 
  \right]
  \mathbf{m}
  \times
  \left(
    \mathbf{p}
    \times
    \mathbf{m}
  \right).
  \label{eq:spin_torque_F1_AHE_symmetric}
\end{equation}
\begin{equation}
\begin{split}
  \frac{{\rm d}\mathbf{p}}{{\rm d} t}
  =
  \frac{\gamma_{0} \hbar}{2e\mu_{0} M_{\rm s}d}
  \frac{g^{*} g_{\rm r} (\beta-\zeta) \tanh[d/(2 \ell_{\rm sf})] \sigma_{\rm AH} E_{x}}
    {g_{\rm sd}^{\prime}(g_{\rm r}+g^{*})}
  \left[
    \frac{p_{y} \lambda \mathbf{m}\cdot\mathbf{p} - m_{y}}{1 - \lambda^{2} (\mathbf{m}\cdot\mathbf{p})^{2}} 
  \right]
  \mathbf{p}
  \times
  \left(
    \mathbf{m}
    \times
    \mathbf{p}
  \right).
  \label{eq:spin_torque_F2_AHE_symmetric}
\end{split}
\end{equation}


\section{Linearized LLG equation}
\label{app:perpswitch}

Linearizing the LLG equation, Eq.~(\ref{eq:LLG_macrospin_AHE}) gives 
\begin{equation}
  \frac{1}{\gamma_{0}}
  \frac{{\rm d}}{{\rm d}t}
  \begin{pmatrix}
    m_{x} \\
    m_{y} 
  \end{pmatrix}
  +
  \mathsf{C}
  \begin{pmatrix}
    m_{x} \\
    m_{y}
  \end{pmatrix}
=
  \frac{\hbar \tanh[d_{2}/(2\ell_{\rm sf}^{\rm F_{2}})]E_{x}}{2e \mu_{0} M_{\rm s}d_{1}}
  \frac{g_{\rm F_{2}}^{*}g_{\rm r(F_{1})} \tilde{\sigma}_{E({\rm F}_{2})}}{g_{\rm sd(F_{2})}^{\prime} (g_{\rm r(F_{1})} + g_{\rm F_{2}}^{*})}
  \frac{1}{1 - \lambda_{1} \lambda_{2} p_{z}^{2}}
  \begin{pmatrix}
    p_{x} \\
    p_{y} 
  \end{pmatrix}.
  \label{eq:macrospin_LLG_AHE_linearized}
\end{equation}
The coefficient matrix $\mathsf{C}$ is given by 
\begin{eqnarray}
  \mathsf{C}
  &=&
  \begin{pmatrix}
    \alpha (H_{\rm K}-M_{\rm s}) & (H_{\rm K}-M_{\rm s}) \\
    -(H_{\rm K}-M_{\rm s}) & \alpha (H_{\rm K}-M_{\rm s}) 
  \end{pmatrix}
\nonumber\\
&&+
  \frac{\hbar \tanh[d_{2}/(2\ell_{\rm sf}^{\rm F_{2}})]E_{x}}{2e \mu_{0} M_{\rm s}d_{1}}
  \frac{g_{\rm F_{2}}^{*}g_{\rm r(F_{1})} \tilde{\sigma}_{E({\rm F}_{2})}}{g_{\rm sd(F_{2})}^{\prime} (g_{\rm r(F_{1})} + g_{\rm F_{2}}^{*})}
\ \ \ \ \ 
  \begin{pmatrix}
    \frac{[1-\lambda_{1}\lambda_{2}(p_{z}^{2}+2p_{x}^{2})]p_{z}}{(1-\lambda_{1}\lambda_{2}p_{z}^{2})^{2}} 
  & 
    -\frac{2 \lambda_{1} \lambda_{2} p_{x} p_{y} p_{z}}{(1-\lambda_{1}\lambda_{2}p_{z}^{2})^{2}}
  \\
    -\frac{2 \lambda_{1} \lambda_{2} p_{x} p_{y} p_{z}}{(1-\lambda_{1}\lambda_{2}p_{z}^{2})^{2}}
  & 
    \frac{[1-\lambda_{1}\lambda_{2}(p_{z}^{2}+2p_{y}^{2})]p_{z}}{(1-\lambda_{1}\lambda_{2}p_{z}^{2})^{2}} 
  \end{pmatrix}. 
\end{eqnarray}
The solutions of Eq.~(\ref{eq:macrospin_LLG_AHE_linearized}) can be
expressed as superpositions of $\exp\{\gamma_{0}[\pm {\rm i}
\sqrt{{\det}[\mathsf{C}] - ({\rm Tr}[\mathsf{C}]/2)^{2}} - {\rm
  Tr}[\mathsf{C}]/2]t\}$.  When the real part of the exponent
($\propto -\gamma_{0}{\rm Tr}[\mathsf{C}]t$) is positive (negative),
the amplitude of $m_{x}$ and $m_{y}$ increases (decrease) with time.
Then, we define the critical electric field to excite the
magnetization dynamics by the condition ${\rm Tr}[\mathsf{C}]=0$.  In
terms of the current density $j=\sigma E_{x}$, the critical current
density is given by Eq.~(\ref{eq:critical_current}).

\subsection{Optimum direction of $\mathbf{p}$ to minimize
  Eq. (\ref{eq:critical_current})}
\label{app:optimum}

When the polarizing layer has only the anomalous Hall effect and no
anisotropic magnetoresistance, the critical current,
Eq.~(\ref{eq:critical_current}) becomes
\begin{equation}
  j_{\rm crit}^{\rm AH}
  =
  -\frac{2\alpha e \mu_{0} M_{\rm s} d_{1} (H_{\rm K}-M_{\rm s})}{\hbar\tanh[d_{2}/(2 \ell_{\rm sf}^{\rm F_{2}})]}
\times
  \frac{(1-\lambda_{1}\lambda_{2} p_{z}^{2})^{2} g_{\rm sd(F_{2})}^{\prime} (g_{\rm r(F_{1})} + g_{\rm F_{2}}^{*}) \sigma_{\rm F_{2}}}
    {(\beta_{\rm F_{2}}-\zeta_{\rm F_{2}})(1-\lambda_{1}\lambda_{2}) p_{y}p_{z} g_{\rm F_{2}}^{*} g_{\rm r(F_{1})} \sigma_{\rm AH(F_{2})}}. 
  \label{eq:critical_current_AHE} 
\end{equation}
This is proportional to
\begin{equation}
  j_{\rm crit}^{\rm AH}
  \propto 
  \frac{(1 - \lambda_{1}\lambda_{2} p_{z}^{2})^{2}}{p_{y} p_{z}},
\end{equation}
where $\lambda_{k}$ is independent of the magnetization direction in this case. 
Then, $j_{\rm crit}^{\rm AH}$ is minimized 
when the polar angle $\theta_{\rm fixed}$ is given by 
\begin{equation}
  \theta_{\rm fixed}
  =
  \tan^{-1}
  \left[
    \sqrt{
      \frac{\lambda_{1}\lambda_{2}+2 - \sqrt{(3 \lambda_{1}\lambda_{2} - 2)^{2} + 8 \lambda_{1} \lambda_{2}}}
        {3\lambda_{1}\lambda_{2} - 2 + \sqrt{(3 \lambda_{1}\lambda_{2} - 2)^{2} + 8 \lambda_{1} \lambda_{2}}}
    }
  \right].
\end{equation}
For the parameters shown in Fig. \ref{fig:critical_currents}, 
the optimum angle is estimated to be $\theta_{\rm fixed}=31.6^{\circ}$. 

\end{widetext}


\end{document}